\documentclass[aps,prb,amsmath,amssymb,reprint,superscriptaddress,preprintnumbers,showpacs,intlimits,longbibliography]{revtex4-1}
\usepackage{bm,latexsym,mathrsfs,enumerate,color}
\usepackage[mathcal]{euscript}
\usepackage[breaklinks=true,unicode=true,urlcolor = blue,colorlinks = true,citecolor = blue,linkcolor = blue]{hyperref}
\usepackage{graphicx}
\usepackage{adjustbox}
\usepackage{array,ragged2e}
\usepackage[cal=boondoxo]{mathalfa}
\usepackage{lastpage}
\DeclareGraphicsExtensions{.pdf,.svg,.eps,.ps,.png,.jpg,.jpeg}
\usepackage{todonotes}

\graphicspath{{../figs/}}
\renewcommand{\vec}[1]{\bm{#1}}
\bibpunct{[}{]}{,}{n}{}{}

\begin{document}
%
%
\title{Chaotic Antiferromagnetic Nano-Oscillator driven by Spin-Torque}

\author{Benjamin Wolba}
\affiliation{Institut f\"ur Theoretische Festk\"orperphysik, Karlsruhe Institute of Technology, 76131 Karlsruhe, Germany}

\author{Olena Gomonay}
\affiliation{Institut f{\"u}r Physik, Johannes Gutenberg-Universit{\"a}t Mainz, D-55128 Mainz, Germany}

\author{Volodymyr P. Kravchuk}
\email[Corresponding author: ]{volodymyr.kravchuk@kit.edu}
\affiliation{Institut f\"ur Theoretische Festk\"orperphysik, Karlsruhe Institute of Technology, 76131 Karlsruhe, Germany}
\affiliation{Bogolyubov Institute for Theoretical Physics of National Academy of Sciences of Ukraine, 03680 Kyiv, Ukraine}



%
%
%
%
\begin{abstract}
We theoretically describe the behavior of a terahertz nano-oscillator based on an anisotropic antiferromagnetic dynamical element driven by spin torque. We consider the situation when the polarization of the spin-current is perpendicular to the external magnetic field applied along the anisotropy easy-axis. We determine the domain of the parametric space (field, current) where the oscillator demonstrates chaotic dynamics. Characteristics of the chaotic regimes are analyzed using conventional techniques such as spectra of the Lyapunov exponents. We show that the threshold current of the chaos appearance is particularly low in the vicinity of the spin-flop transition. In this regime, we consider the mechanism of the chaos appearance in detail when the field is fixed and the current density increases. 
We show that the appearance of chaos is preceded by a regime of  quasiperiodic dynamics on the surface of a two-frequency torus arising in phase space as a result of the Neimark-Sacker bifurcation.
\end{abstract}
%

\maketitle


\section{Introduction}

Neuromorphic computing is a rapidly developing field inspired by the idea of emulating biological processes in the brain \cite{Markovic20,Roy19}. Since neurons demonstrate a rhythmic activity, non-linear nano-oscillators of different physical nature \cite{Segall17,Torrejon17,Pickett12} are considered as candidates for artificial neurons. Spin-torque nano-oscillators are of special interest because their dynamical regimes are easy-tunable by means of electrical current. They are used to emulate single neurons \cite{Torrejon17}, as well as small neural networks \cite{Romera18}.

Several concepts of neuromorphic computing rely on stochastic (chaotic) oscillator behavior; examples are reservoir computing \cite{Markovic19,Torrejon17,Jaeger04,Tsunegi19,Choi19a} or spike-based encoding \cite{Roy19,Khymyn18,Zhao15,Matsumoto19}. Thereby, a spin-torque nano-oscillator that demonstrates controllable chaotic dynamics is of high importance for these concepts. The simplest spin-torque nano-oscillator is made of a collinear antiferromagnet (AFM) and is essentially a non-linear dynamical system with a four-dimensional phase space \cite{Gomonay10,Cheng16a}. This feature makes an AFM nano-oscillator a natural candidate for the realization of chaotic dynamic regimes. The possibility of chaos in such systems was recently pointed out in Ref.~\onlinecite{Parthasarathy19a}. An advantage of AFM nano-oscillators is their fast dynamics in the THz regime. This may allow for the construction of ultrafast artificial neurons \cite{Khymyn18}.

Besides neuromorphic computing, spin-torque nano-oscillators are widely considered as building blocks for various spintronic devices, e.g. memory elements, field detectors, or microwave generators \cite{Locatelli13a,Chen16a}.

Here, we provide a general analysis of possible dynamical regimes of an AFM nano-oscillator driven by a spin-current. Two external parameters control the system, namely the applied magnetic field and the strength of the current. The dynamical regimes were studied in a large domain of the 2D parametric space, which is of interest for spintronic applications. First, we formulate the model and derive the basic equation of motion, see Sec.~\ref{sec:model} supplemented with App.~\ref{app:eqs}. Next, we consider possible fixed points of the system and study their stability, see Sec.~\ref{sec:statics} and App.~\ref{app:stability}. Then we analyze the regimes in phase space where stable limit-cycles are present, and we provide analytical expressions for the parameters of the limit cycles and analyze their stability, see Sec.~\ref{sec:LC} and App.~\ref{app:Monodromy}. In the last Sec.~\ref{sec:chaos} we demonstrate the possibility of chaos and hyperchaos in the system and analyze characteristics of the chaotic dynamics within parameter space. Also, we identify the mechanism of the chaos appearance in the vicinity of the spin-flop transition.
 
\section{Model}\label{sec:model}
We consider a two-sublattice antiferromagnetic (AFM) film. The magnetization of each of the sublattices is modeled by means of a continuous function $\vec{M}_i=\vec{M}_i(\vec{r},t)$ of constant amplitude $|\vec{M}_{i}|=M_s$, where $i=1,2$. In the following, it is instructive to introduce the unit magnetization vector $\vec{\mu}_i=\vec{M}_i/M_s$ together with the ferro- $\vec{m}=(\vec{\mu}_1+\vec{\mu}_2)/2$ and antiferromagnetic $\vec{n}=(\vec{\mu}_1-\vec{\mu}_2)/2$ vector order parameters. Note that $\vec{m}\cdot\vec{n}=0$ and $\vec{m}^2+\vec{n}^2=1$.  

The magnetization dynamics is described by the set of Landau-Lifshitz equations
\begin{equation}\label{eq:LL}
	\begin{split}
	\partial_t\vec{\mu}_i=&\frac{\gamma}{M_s}\left[\vec{\mu}_i\times\frac{\delta E}{\delta\vec{\mu}_i}\right]+\alpha\left[\vec{\mu}_i\times\partial_t\vec{\mu}_i\right]+\\
	&+\sigma\left[\vec{\mu}_i\times\vec{p}\times\vec{\mu}_i\right],\qquad i=1,2
	\end{split}
\end{equation}
supplemented with the Gilbert damping and the spin-torque term in the form proposed by Slonczewski \cite{Slonczewski96,Slonczewski02,Xiao04}. Here, $\gamma>0$ is the gyromagnetic ratio, and $\alpha$ is the damping constant. The strength of the spin torque is determined by the constant $\sigma=\gamma\hbar PJ/(2|e|LM_s)$, where $\hbar$ denotes the Planck constant, $e$ is the electron charge, $L$ is the thickness of the AFM film, $J>0$ and $0<P<1$ are the density and the rate of the spin-polarization of the current which flows perpendicularly to the film. The geometry of the studied system is represented by a pillar structure that consists of three layers: ferromagnetic polarizer, nonmagnetic spacer, and the studied AFM free layer (see Fig.~\ref{fig:static}(a,b)). The unit vector $\vec{p}$ indicates the direction of the spin polarization. 

Two equations of the set \eqref{eq:LL} are coupled by means of the energy functional $E=E[\vec{\mu}_1,\vec{\mu}_2]$ which includes intra- and inter-lattices exchange interactions \eqref{eq:Eex}, uniaxial anisotropy with the easy-axis oriented perpendicularly to the AFM film \eqref{eq:Ean} and interaction with an external magnetic field $\vec{B}$ \eqref{eq:Ezee}. In the following, we consider the case when the applied field and the anisotropy field $B_{an}$ are much smaller compared to the exchange field $B_{ex}$ acting between the sublattices. The fields $B_{ex}$ and $B_{an}$ are introduced in \eqref{eq:Eex} and \eqref{eq:Ean}, respectively. In this limit, the dynamics of the N{\'e}el vector $\vec{n}$ is governed by the Lagrangian \cite{Baryakhtar80,Gomonay10}
\begin{subequations}\label{eq:Lagr}
	\begin{align}
	&\mathcal{L}=\left[\dot{\vec{n}}-(\vec{b}\times\vec{n})\right]^2-\mathcal{W},\\
	& \mathcal{W}=\partial_i\vec{n}\cdot\partial_i\vec{n}-(\vec{n}\cdot\hat{\vec{z}})^2
	\end{align}
supplemented with the Rayleigh dissipation function of density \cite{Gomonay10}
\begin{align}\label{eq:R}
	\mathcal{R}=\bar{\alpha}\dot{\vec{n}}^2-j[\vec{n}\times\vec{p}]\cdot\dot{\vec{n}},
\end{align}
\end{subequations}
where the constraint $|\vec{n}|=1$ is presumed. Thus, the dynamical system under consideration is controlled by three parameters: the dimensionless magnetic field  $\vec{b}=\vec{B}/B_{sf}$ in units of the spin-flop field $B_{sf}=\sqrt{B_{ex}B_{an}}$, the dimensionless current density $j=PJ/J_0$ with $J_0=B_{an}|e|LM_s/\hbar$, and the modified damping coefficient $\bar{\alpha}=\alpha\sqrt{B_{ex}/B_{an}}$. In \eqref{eq:Lagr} we utilized the dimensionless space-time coordinates $\vec{\varrho}=\vec{r}/\ell$ and $\tau=t\omega_{0}$, where $\ell=\sqrt{A/(B_{an}M_s)}$ is the exchange length and $\omega_0=\gamma B_{sf}$ is the frequency of the uniform AFM resonance. Here, $A$ is the constant of the nonunifom exchange interaction introduced in \eqref{eq:Eex-mn}. This dimensionless system of units was used throughout the following part of the main text.

Having determined the N{\'e}el vector $\vec{n}$ by means of the theory \eqref{eq:Lagr}, one obtains the magnetization vector in the form \cite{Baryakhtar80,Gomonay10}
\begin{equation}\label{eq:m}
	\vec{m}\approx\sqrt{B_{an}/B_{ex}}\left(\dot{\vec{n}}+\vec{n}\times\vec{b}\right)\times\vec{n}.
\end{equation}

In the following, we focus on the case $\vec{p}=-\hat{\vec{y}}$ and $\vec{b}=b\hat{\vec{z}}$. The spatially uniform solutions $\vec{n}=\vec{n}(\tau)$ of \eqref{eq:Lagr} are determined by the equation
\begin{equation}\label{eq:n-dyn}
	\begin{split}
\Bigl[\ddot{\vec{n}}+2b(\dot{\vec{n}}\times\hat{\vec{z}})&+(\hat{\vec{z}}\cdot\vec{n})(b^2-1)\hat{\vec{z}}\\
&+\bar{\alpha}\dot{\vec{n}}+\frac{j}{2}(\vec{n}\times\hat{\vec{y}})\Bigr]\times\vec{n}=0	
\end{split}
\end{equation}
It is convenient to enforce the constraint $|\vec{n}|=1$ using the spherical parameterization $\vec{n}=\sin\theta\left(\hat{\vec{x}}\cos\phi+\hat{\vec{y}}\sin\phi\right)+\cos\theta\hat{\vec{z}}$. In this case, Eq.~\eqref{eq:n-dyn} reduces to 
\begin{subequations}\label{eq:angles}
	\begin{align}
\label{eq:theta}	&\ddot{\theta}+\bar{\alpha}\dot{\theta}+\sin\theta\cos\theta\left[1-(\dot{\phi}-b)^2\right]=\frac{j}{2}\cos\phi,\\
\label{eq:phi}	&\ddot{\phi}+2\cot\theta\dot{\theta}(\dot{\phi}-b)+\bar{\alpha}\dot{\phi}=-\frac{j}{2}\cot\theta\sin\phi.
	\end{align}
\end{subequations}
The set of nonlinear equations \eqref{eq:angles} is the main subject of this paper.

\section{Static equilibrium states}\label{sec:statics}

\begin{figure}
	\includegraphics[width=\columnwidth]{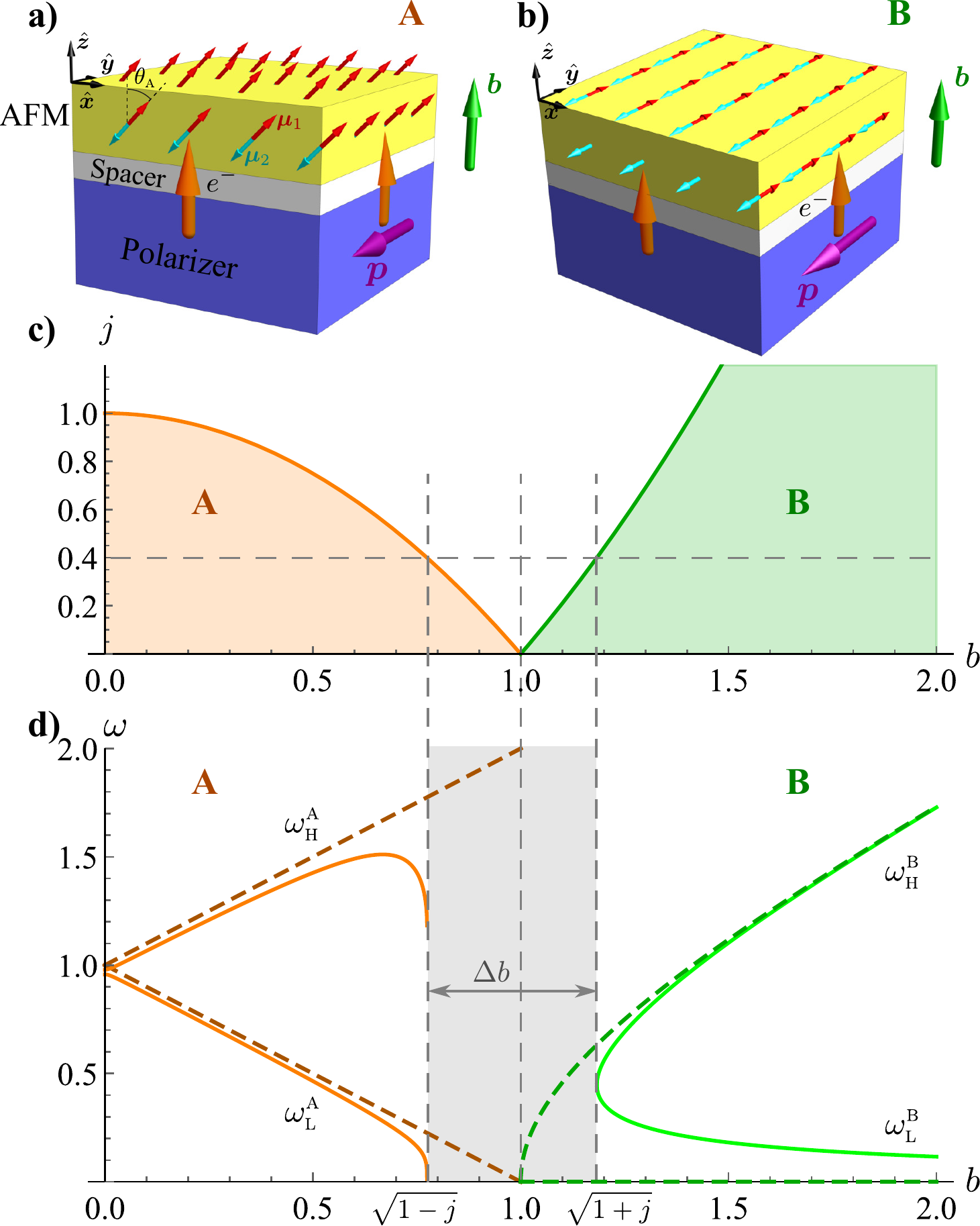}
	\caption{Geometry of the studied setup and properties of equilibrium states. The two possible stable static solutions `A' (represented by \eqref{eq:A}) and `B' (represented by \eqref{eq:B}) are shown on insets a) and b), respectively. c) -- stability domains of solutions `A' and `B'. d) -- eigenfrequencies of the solutions `A' and `B' for the cases $j=0$ (dashed lines) and $j=0.4$ (solid lines). The shadowed region of width $\Delta b\approx j$ indicates a gap where stable static solutions cannot exist. Insets c) and d) are made for $\bar{\alpha}=0$.}\label{fig:static}
\end{figure}

Let us start with static equilibrium solutions. For the case $j=0$, the considered AFM system is well studied~\cite{Akhiezer68}. It is known that depending on the applied magnetic field, there are two equilibrium states: one state with uniform polarization along the anisotropy axis $\vec{n}||\hat{\vec{z}}$ for $0<|b|<1$ (solution A), and the spin-flop state $\vec{n}\perp\hat{\vec{z}}$ when $|b|>1$ (solution B). In the first and latter cases the equilibrium orientation of the N{\'e}el vector is doubly and continuously degenerate, respectively. Linear eigenexcitations of both solutions are characterized by two frequency modes. The high- and low-frequency modes of solution (A) are, $\omega_{\textsc{h}}^{\textsc{a}}=1+b$ and $\omega_{\textsc{l}}^{\textsc{a}}=1-b$, respectively. They are plotted as dashed orange lines in Fig.~\ref{fig:static}d). Softening of the low-frequency mode for $b=1$ corresponds to the instability of solution (A) which results in a transition to the spin-flop state (B). The high- and low-frequency modes of the solution (B) are, $\omega_{\textsc{h}}^{\textsc{b}}=\sqrt{b^2-1}$ and $\omega_{\textsc{l}}^{\textsc{b}}=0$, respectively.  They are plotted as dashed green lines in Fig.~\ref{fig:static}d). Zeroing of the low-frequency mode reflects the continuous degeneracy of the solution B, i.e. all orientations of the N{\'e}el vector within $x-y$-plane are energetically equivalent.

Applying a spin-torque changes the properties of both states (A) and (B). For state (A), the magnetizations of the sublattices take a turn in the plane perpendicular to $\vec{p}$ such that the N{\'e}el order parameter $\vec{n}$ makes an angle $0<\theta_{\textsc{A}}<\pi/4$ with the anisotropy axis, see Fig.~\ref{fig:static}a). Here,
\begin{equation}\label{eq:A}
	\sin2\theta_{\textsc{A}}=\frac{j}{j_c(b)},\qquad \phi_{\textsc{A}}=0
\end{equation}
with $j_c=1-b^2$ satisfies static form of Eqs.~\eqref{eq:angles}.
Note that one more solution can be obtained from \eqref{eq:A} by means of the transformation $\theta\to\pi-\theta$, $\phi\to\phi+\pi$ (equivalently $\vec{n}\to-\vec{n}$) that also satisfies \eqref{eq:angles} and has the same properties as \eqref{eq:A}. State (A) is stable within the parameter domain $b<1$ and $j<1-b^2$, see Fig.~\ref{fig:static}c). The spin torque removes the degeneracy of the high- and low-frequency modes at zero magnetic fields, resulting in the splitting $\propto j^2$. Namely, $\omega_{\textsc{h}}^{\textsc{a}}=1-\frac18j^2$ and $\omega_{\textsc{l}}^{\textsc{a}}=1-\frac14j^2$ if $b=0$. The spin torque also reduces the critical field $b_{c1}=\sqrt{1-j}$ when the state (A) becomes unstable. In the vicinity of the instability the high- and low-frequency modes demonstrate the following asymptotic behavior $\omega_{\textsc{h}}^{\textsc{a}}\approx c_1+\tilde{c}_1\sqrt{b_{c1}-b}$, and $\omega_{\textsc{l}}^{\textsc{a}}\approx c_2\sqrt[4]{b_{c1}-b}$, respectively. Here, $c_1=\sqrt{(1+3b_{c1}^2)/2}$ and  $c_2=b_{c1}^{1/4}(1-b_{c1}^2)^{3/4}/c_1$.

The static state \eqref{eq:A} has the magnetization
\begin{equation}\label{eq:mA}
	\vec{m}_{\textsc{a}}=\sqrt{\frac{B_{an}}{B_{ex}}}\frac{b}{2}\left[-\frac{j}{j_c}\hat{\vec{x}}+\left(1-\sqrt{1-\frac{j^2}{j_c^2}}\right)\hat{\vec{z}}\right].
\end{equation}

\begin{figure*}
	\includegraphics[width=\textwidth]{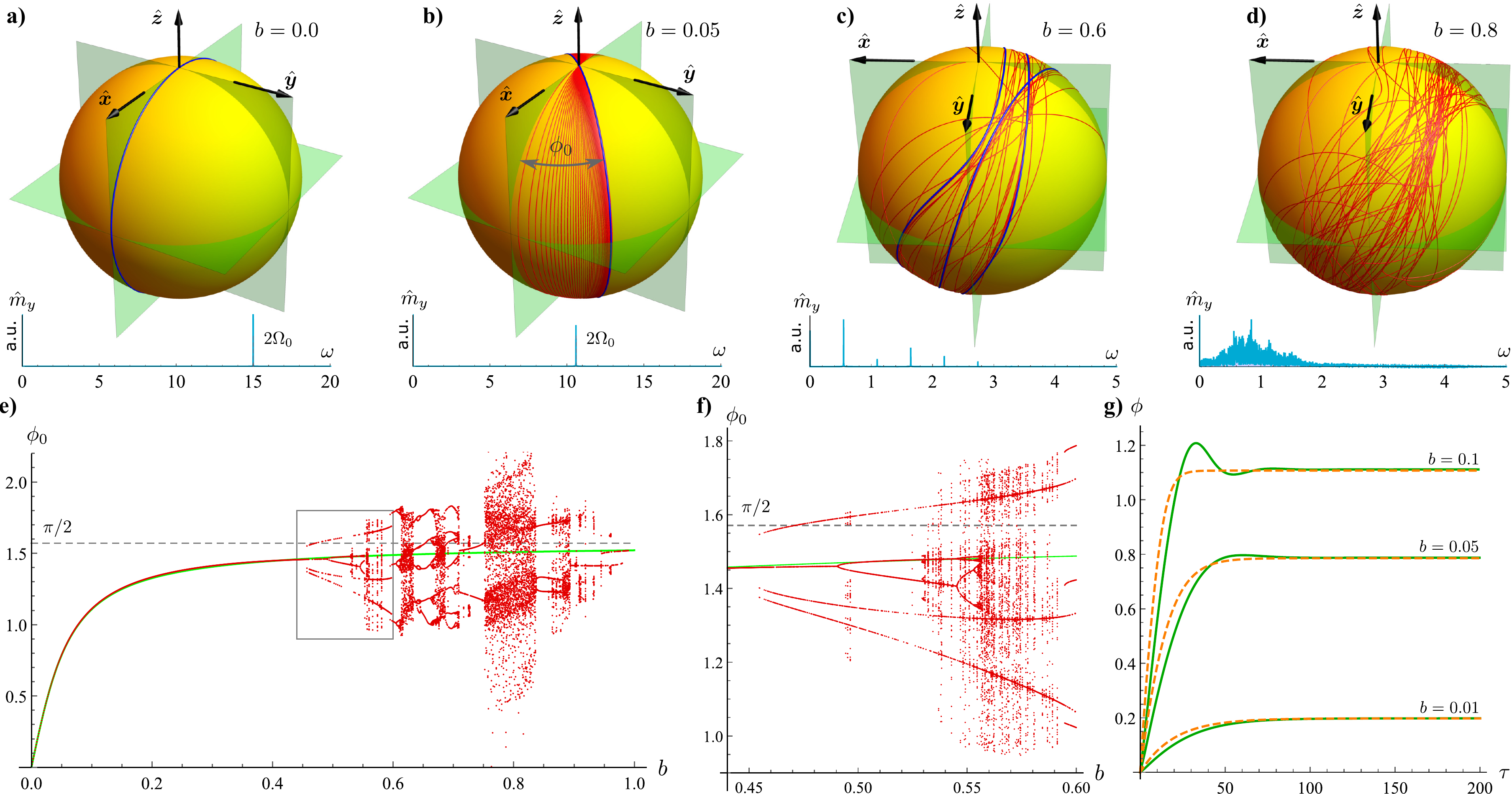}
	\caption{Evolution of dynamical regimes of the oscillator with increasing magnetic field. The upper row demonstrates three different types of dynamics in terms of trajectories of the N{\'e}el vector $\vec{n}$ (thin red lines) on the surface of a unit sphere. The thick blue line shows the limit cycle (if it exists). The trajectories are obtained as the numerical solution of Eqs.~\eqref{eq:angles} with parameters $j=1.5$ and $\bar{\alpha}=0.1$ in the time domain $0<\tau<\tau_{\text{max}}$ with $\tau_{\text{max}}=10^3$, and for the initial conditions $\phi(0)=0$, $\phi'(0)=0$, $\theta(0)=0.1$, $\theta'(0)=0$. The Fourier transform $\hat{m}_y(\omega)$ of a magnetization component \eqref{eq:m} enables us to distinguish regular dynamics (a-c) from the chaotic regime (d). Vertical dashed ticks show the frequency $\omega=2\Omega_0$. The azimuthal coordinates of the intersections of the trajectories with the horizontal half-plane $\{z=0,\,y>0\}$ in the time interval $0.8\tau_{\text{max}}<\tau<\tau_{\text{max}}$ are denoted as $\phi_0$. The coordinates $\phi_0$ collected for all values of the magnetic field compose the bifurcation diagram shown in panel (e). Panel (f) shows the zoomed boxed region of panel (e). The solid green line shows the approximation $\phi_0\approx\arctan(2b/\bar{\alpha})$. Panel (g) compares the numerically obtained dependence $\phi(\tau)$ (green solid lines) with the approximation \eqref{eq:phi-sol} (dashed orange lines) in the limit of small fields. }\label{fig:dynamics}
\end{figure*}

Note that Eqs.~\eqref{eq:angles} allow one more static solution (A'), which one obtains from \eqref{eq:A} by means of the transformation $\theta_{\textsc{a}'}=\theta_{\textsc{a}}+\pi/2$, $\phi_{\textsc{a}'}=\phi_{\textsc{a}}+\pi$. It is unstable for all values of $b$ and $j$ (see App.~\ref{app:stability} for details) and is not considered in the main text. 

The spin-torque removes the continuous degeneracy of the state (B) and leads the sublattices magnetization to orient itself parallel to the direction of the spin polarization: $\vec{n}=\pm\vec{p}$. In terms of the angular parameterization 
\begin{equation}\label{eq:B}
	\theta_{\textsc{b}}=\pi/2,\qquad \phi_{\textsc{b}}=\pm\pi/2,
\end{equation}
see Fig.~\ref{fig:static}b).
Solution \eqref{eq:B} is stable within the parameter domain $b>1$ and $j<|j_c|=b^2-1$. High- and low-frequency modes of the state (B) are $\omega_{\textsc{h}}^{\textsc{b}}=\sqrt{|j_c|+\sqrt{j_c^2-j^2}}/\sqrt{2}$ and $\omega_{\textsc{l}}^{\textsc{b}}=\sqrt{|j_c|-\sqrt{j_c^2-j^2}}/\sqrt{2}$, respectively. See Fig.~\ref{fig:static}d). The spin-torque increases the minimal magnetic field $b_{c2}=\sqrt{1+j}$ above which the state (B) is stable. So, a small spin-torque opens a gap $b_{c1}<b<b_{c2}$ in the vicinity of the spin-flop field, where no static solitons can exist. As it will be shown latter, chaos appears mostly in this gap.

The magnetization of the static state (B) is $\vec{m}_{\textsc{b}}=\sqrt{B_{an}/B_{ex}}b\hat{\vec{z}}$.

The analysis provided in this section was made assuming vanishing damping. Details of the stability analysis for possible static states, including the damping effects, are provided in Appendix~\ref{app:stability}.

\section{Regular dynamics and transition to chaos}\label{sec:LC}

Let us start with the simple case $b=0$. Increasing the current from $j=0$ to $j=1$, we obtain the static stable solution (A) shown in Fig.~\ref{fig:static}(a) whose inclination angle varies from $\theta_{\textsc{a}}=0$ to $\theta_{\textsc{a}}=\pi/4$. With a further increase in the current $j>1$, equations \eqref{eq:angles} do not allow stable static solutions any more. However, in the particular case $b=0$, Eq.~\eqref{eq:phi} is satisfied if $\phi=0$. This means that the dynamics of the N{\'e}el vector $\vec{n}$ is constrained within $z-x$ plane. The orientation of the vector $\vec{n}$ is determined by the polar angle $\theta(\tau)$ which satisfies the equation of motion of a driven nonlinear pendulum with damping
\begin{equation}\label{eq:pend}	\ddot{\theta}+\bar{\alpha}\dot{\theta}+\sin\theta\cos\theta=j/2.
\end{equation}  
The dynamics of the antiferromagnetic oscillator for this case was analyzed in detail in Ref.~\onlinecite{Parthasarathy19a} and will not be considered here. We only note that for large currents $j\gg1$ the dynamics of $\vec{n}$ takes the form of a quasi-uniform rotation $\theta\approx\Omega\tau$ with $\Omega=j/(2\bar{\alpha})$. On a unit sphere, this dynamics is represented by a circular limit cycle in $x-z$-plane, see Fig.~\ref{fig:dynamics}a.

Now we introduce a small magnetic field $b\ll1$. This limit allows us to assume that $\dot{\phi}\ll\dot{\theta}$. I.e. the system has two well distinguished time scales, namely, vector $\vec{n}$ rapidly rotates with the angular velocity $\dot{\theta}$ within a vertical plane, which slowly rotates about the $z$-axis with the angular velocity $\dot\phi$. This kind of dynamics is shown in Fig.~\ref{fig:dynamics}b. The corresponding approximate analytical solution can be derived from \eqref{eq:angles} in the following way: Assuming that the current is large we conclude from \eqref{eq:theta} that $\theta\approx\Omega\tau$ with $\Omega=j\cos\phi/(2\bar{\alpha})$. Substituting this approximation into \eqref{eq:phi} and neglecting terms involving $\ddot{\phi}$ and $\bar{\alpha}\dot{\phi}$ we obtain a solution for $\phi(\tau)$ in the implicit form
\begin{equation}\label{eq:phi-sol}
	\phi-\eta\ln|\cos\phi-\eta\sin\phi|\approx b(1+\eta^2)t,\qquad \eta=\frac{\bar{\alpha}}{2b}.
\end{equation}
The solution \eqref{eq:phi-sol} satisfies the initial condition $\phi(0)=0$, which corresponds to the static state \eqref{eq:A}. Thus, if one starts with the static state (A) and increases the current step-by-step such that $j>j_c$, then the resulting dynamics can be approximated by \eqref{eq:phi-sol}. Indeed, as it can be seen in Fig.~\ref{fig:dynamics}g, in the limit of small fields, the approximation \eqref{eq:phi-sol} demonstrates a good agreement with the numerical solution of \eqref{eq:angles}.

As it follows from \eqref{eq:phi-sol}, for $\tau\to\infty$ the orientation of the plane of rotation of $\vec{n}$ approaches the asymptotic value $\phi\to\phi_0=\arctan(1/\eta)$. The corresponding limit cycle on the unit sphere has the following characteristics $\phi=\phi_0$, $\theta=\Omega_0\tau$, where 
\begin{equation}\label{eq:lim-cycle}
	\phi_0=\arctan(2b/\bar{\alpha}),\qquad \Omega_0=\frac{j}{2\sqrt{\bar{\alpha}^2+4b^2}}.
\end{equation} 
The typical time of approaching the limit cycle is $\Delta\tau\approx2\bar{\alpha}/(\bar{\alpha}^2+4b^2)$. Note that for larger fields, the plane of rotation of vector $\vec{n}$ approaches its asymptotic orientation $\phi_0$ in an oscillatory manner, see solid lines Fig.~\ref{fig:dynamics}g. This behavior is not described by the approximation \eqref{eq:phi-sol} because terms involving $\ddot{\phi}$ in \eqref{eq:phi} were neglected during the derivation of \eqref{eq:phi-sol}. Despite this, the asymptotic value $\phi_0$ is valid for relatively large fields, see Fig.~\ref{fig:dynamics}e. 

Interestingly, in the particular case $b=1$, the limit cycle \eqref{eq:lim-cycle} is an \emph{exact} solution of Eqs.~\eqref{eq:angles}. Using the monodromy matrix technique (see App.~\ref{app:Monodromy}) we found that this solution demonstrates a couterintuitive stability up to relatively high currents.


Experimentally, the described dynamics can be detected by observing the oscillations of the magnetization vector $\vec{m}$, which are measurable in contrast to $\vec{n}$. Using \eqref{eq:m} one easily obtains that for the limit cycle \eqref{eq:lim-cycle} the magnetization vector reads
\begin{equation}\label{eq:m-lim-cycle}
	\vec{m}\approx\sqrt{\frac{B_{an}}{B_{ex}}}\left[\Omega_0(\vec{\varepsilon}\times\hat{\vec{z}})-\vec{\varepsilon}\frac{b}{2}\sin2\Omega_0\tau+\hat{\vec{z}}b\sin^2\Omega_0\tau\right],
\end{equation}
where $\vec{\varepsilon}=\hat{\vec{x}}\cos\phi_0+\hat{\vec{y}}\sin\phi_0$. Thus, the presence of the doubled frequency $2\Omega_0$ in the spectra of electromagnetic radiation of the nano-oscillator indicates the presence of the limit cycle \eqref{eq:lim-cycle}. This is reflected by the Fourier spectra in Fig.~\ref{fig:dynamics}(a,b). Although the oscillatory part of the magnetization vector \eqref{eq:m} vanishes for $b=0$, the spectral line at $2\Omega_0$ is still present, see Fig.~\ref{fig:dynamics}a. This is because the actual solution $\theta(\tau)$ of the pendulum equation Eq.~\eqref{eq:pend} deviates from the uniform rotation $\theta\approx\Omega_0\tau$.

As it follows from \eqref{eq:lim-cycle}, the frequency $\Omega_0$ decreases with the field and so  $\dot{\theta}$ also decreases. This results in a violation of the assumption about the separation of time scales $\dot{\theta}\gg\dot\phi$. It leads to a significant deviation of the trajectories on the unit sphere from the approximation \eqref{eq:phi-sol}. The resulting dynamics can be still regular (Fig.~\ref{fig:dynamics}c) or chaotic (Fig.~\ref{fig:dynamics}d). The appearance of chaos drastically changes the spectra of the magnetization, see Fig.~\ref{fig:dynamics}d, and therefore can be experimentally detected using spectral analysis of the radiation emitted by the oscillator. 

To track the appearance of chaos in our system, we employed a method similar to the use of Poincar{\'e} sections: We consider the intersection points of the trajectories on the unit sphere with the horizontal half-plane $\{z=0,\,y>0\}$ for times $\tau>\tau^*$ where time $\tau^*$ is much larger than all typical timescales of the system. The regular dynamics is characterized by a constant number of intersection points for all times, e.g. one point for Figs.~\ref{fig:dynamics}a,b and three points for Fig.~\ref{fig:dynamics}c. For chaotic or quasiperiodic behavior, the number of intersection points increases in time and becomes quite large for long times of integration of \eqref{eq:angles}, see Fig.~\ref{fig:dynamics}d. Since an intersection point lies on the circle of unit radius, it is characterized by a single polar coordinate $\phi$. For the circular limit cycles described above $\phi=\phi_0$. So we use this notation for the intersection point coordinates even if the limit circle is more complicated (Fig.~\ref{fig:dynamics}c) or it does not exist (Fig.~\ref{fig:dynamics}d). Plotting all coordinates $\phi_0$ obtained for a range of magnetic fields, we build a typical bifurcation diagram shown in Figs.~\ref{fig:dynamics}e,f. As one can see, chaotic regimes are intercepted by windows of regular dynamics characterized by period multiplication of the limit cycles. Note that this technique does not allow one to distinguish between chaotic and quasiperiodic dynamics. However, as it will be shown in the next section, the domains of quasiperiodicity are quite narrow.

\section{Characteristics of the chaotic regime}\label{sec:chaos}
\begin{figure}
	\includegraphics[width=\columnwidth]{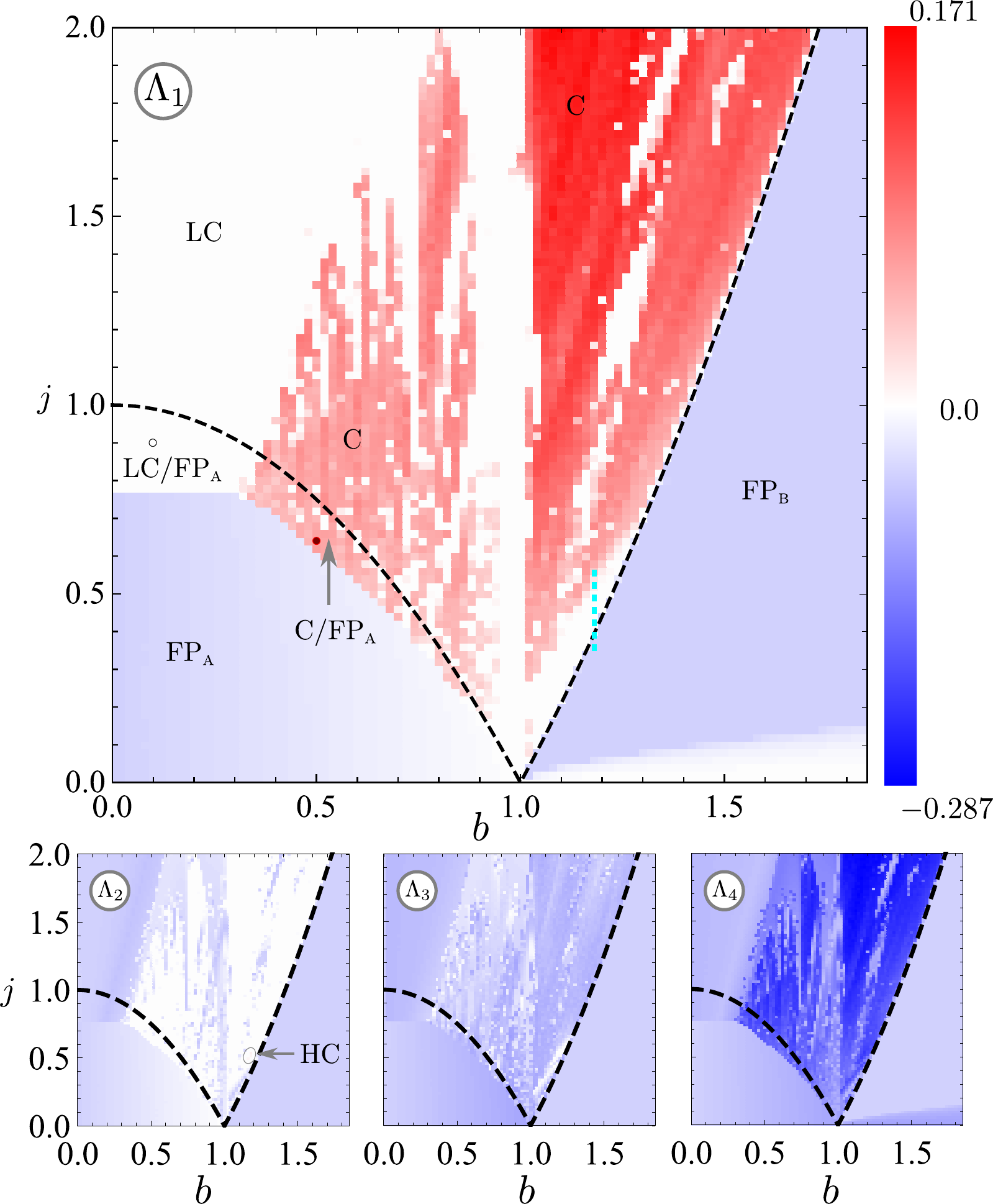}
	\caption{Maps for the Lyapunov exponents $\Lambda_1\ge\Lambda_2\ge\Lambda_3\ge\Lambda_4$ in parameter space obtained for $\alpha=0.1$ and initial conditions \eqref{eq:ini}. The abbreviations denote different types of attractors realized in the phase space: FP$_\textsc{a}$ and FP$_\textsc{b}$ ($\Lambda_1<0$) -- fixed points A and B, respectively (see Fig.~\ref{fig:static}ab), LC ($\Lambda_1=0$) -- stable limit cycle, C ($\Lambda_1>0$) -- chaotic strange attractor. For some regions of the parameter space, two attractors can coexist. HC denotes a small area of the parameter space, where hyper-chaos is possible ($\Lambda_2>0$). The dashed line shows the boundary of stability domains $j=|1-b^2|$ of the static solutions A and B, see Fig.~\ref{fig:static}c. The notations LC/FP$_\textsc{a}$ and C/FP$_\textsc{a}$ specify areas, where a fixed point of type A coexists in the phase space with a limit cycle and a strange attractor, respectively. The corresponding examples of the attraction basins are shown in Fig.~\ref{fig:basins} a) and b) for the open and filled points, respectively. }\label{fig:Lambda1}
\end{figure}

Spectra of the Lyapunov exponents (LE) are a universal tool for studying the complicated dynamics of ergodic systems \cite{ArkadyPikovsky16,Cencini09}. In order to utilize this technique we consider our system \eqref{eq:angles} as a flow $\dot{\vec{x}}=\vec{f}(\vec{x})$, with $\vec{x}=(\theta,\dot{\theta},\phi,\dot{\phi})\in\mathbb{R}^4$ being a time-dependent vector of 4-dimensional phase space. The phase space is equipped with the Euclidean metric and norm $||\vec{x}||=\sqrt{\sum_{i=1}^4x_i^2}$. The latter enables us to introduce the distance $\Delta(\tau)=||\vec{x}(\tau)-\tilde{\vec{x}}(\tau)||$ between two trajectories $\vec{x}$ and $\tilde{\vec{x}}$. The nature of the long-term evolution of the distance $\Delta(\tau)$ for the case $\Delta(0)\to0$ characterizes the instability of the trajectories with respect to the initial conditions. The rate of divergence of the trajectories is determined by the largest LE \cite{Cencini09}
\begin{equation}\label{eq:Lambda1}
	\Lambda_1=\lim\limits_{\begin{smallmatrix}\tau\to\infty\\\Delta(0)\to0
	\end{smallmatrix}}\frac{1}{\tau}\ln\frac{\Delta(\tau)}{\Delta(0)},
\end{equation}
which is shown in Fig.~\ref{fig:Lambda1}. The value of $\Lambda_1$ enables one to deduce a general features of the dynamics, e.g. for the case $\Lambda_1>0$ the system dynamics is chaotic. However, the complete characteristics of the dynamics of a dissipative system is determined by the types of attractors present in phase space. An attractor can be identified from the full spectrum of the LEs whose number coincides with the dimension $d=4$ of the phase space \footnote{The Lyapunov exponents have an intuitive geometrical meaning: under the time evolution an infinitesimally small $d$-dimensional sphere in the phase space is deforming into an ellipsoid with semi-axes $\propto e^{\Lambda_i\tau}$.}. The key instrument here is Oseledets multiplicative ergodic theorem \cite{Oseledets68} which guaranties that all trajectories belonging to one attractor have the same LE spectrum. This means that any trajectory can be chosen for determining the LEs, in the other words, the LEs are independent on the initial conditions. Using the algorithm proposed by Benettin at al. \cite{Benettin80,Benettin80a} we calculated all four LEs for a wide range of parameters $b$ and $j$, see Fig.~\ref{fig:Lambda1}. For all points of the studied parameter space the sum of all LEs is negative $S_d=\sum_{i=1}^d\Lambda_i=-2\bar{\alpha}$ \footnote{The statement $S_d=-2\bar{\alpha}$ was established numerically for a large range of the parameters $b$, $j$,  $\bar{\alpha}$ and initial conditions.}. This reflects the dissipative nature of our system, since negative $S_d=\langle\vec{\nabla}\cdot\vec{f}(\vec{x})\rangle_\tau$ indicates shrinking volumes in phase space \cite{Cencini09}. Thus, we can characterize the dynamics of the system by the type of attractors formed in phase space. 

\begin{figure}
	\includegraphics[width=\columnwidth]{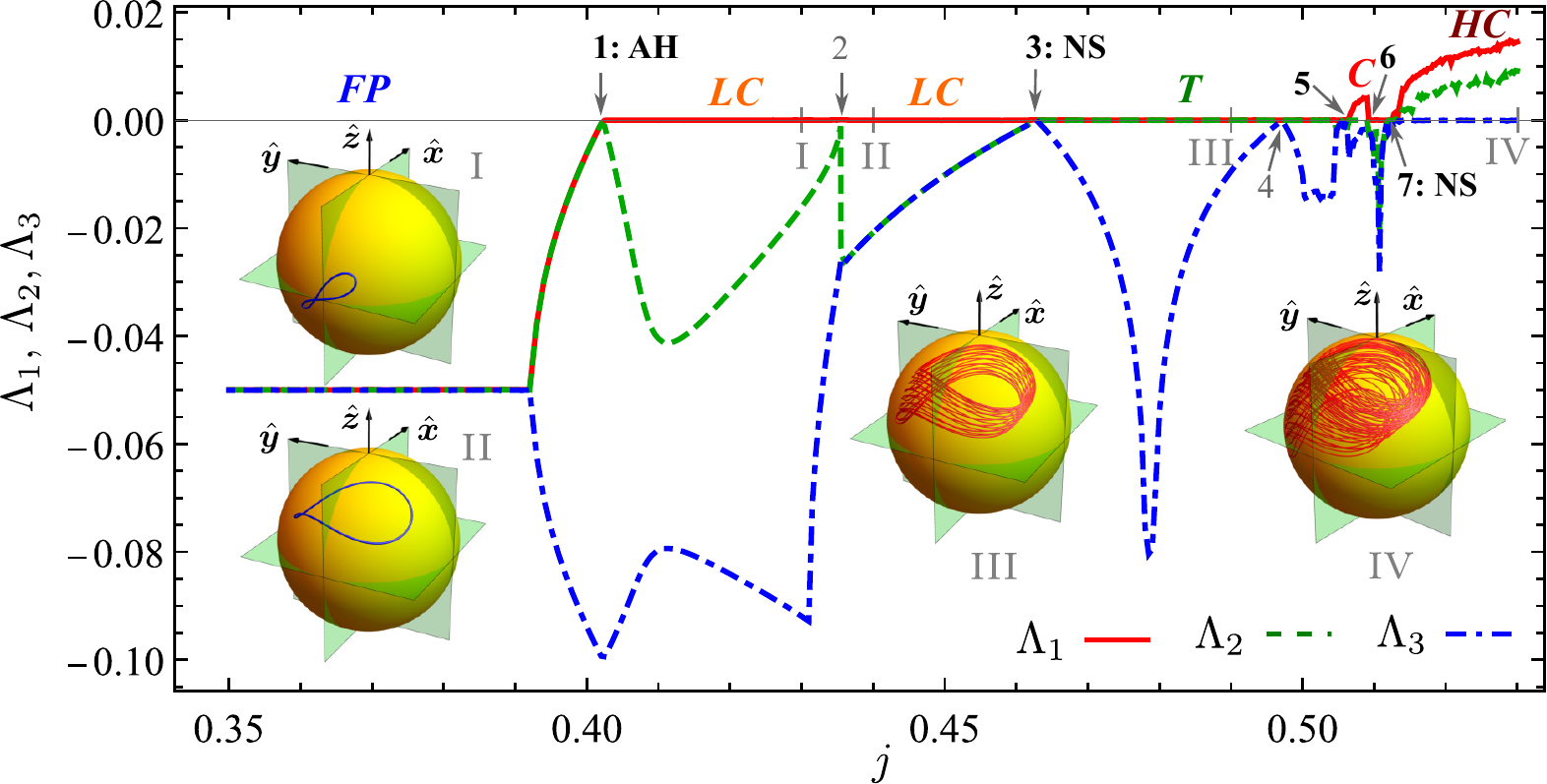}
	\caption{The transition to chaos along line $\{b=1.18,0.35\le j\le 0.53$ (see Fig.~\ref{fig:Lambda1}) is explored by analyzing the variation of the largest three Lyapunov exponents. The exponent $\Lambda_4$ is always negative and, therefore, is not shown here. The bifurcations are numbered from 1 to 7. Bifurcations 2 and 4 do no result in a change of the attractor type; they are shown in gray. Abbreviations AH and NS denote Andronov-Hopf and Neimark-Sacker bifurcations, respectively. The following types of dynamics are identified: fixed point (FP) -- to the left from point 1; periodical dynamics along a limit cycle (LC) -- between points 1, 3 and 6, 7; quasiperiodic dynamics on a 2-frequency torus (T) -- between 3, 5; chaos (C) -- between 5, 6; hyperchaos (HC) -- to the right from point 7. Insets show examples of the limit cycles (I,II), as well as $\vec{n}$-vector trajectories on the surface of the unit sphere for quasiperiodic (III) and chaotic (IV) types of dynamics. The parameters are the same as for Fig.~\ref{fig:Lambda1}.}\label{fig:transition}
\end{figure}

The signs of the ordered LEs compose the signature of the attractor. Using the obtained maps of LEs we distinguish the following signatures: $\langle-,-,-,-\rangle$ -- stable fixed point corresponding to a static equilibrium state, $\langle0,-,-,-\rangle$ -- stable limit cycle corresponding to periodical dynamics, $\langle+,0,-,-\rangle$ -- chaos, $\langle+,+,0,-\rangle$ -- hyper-chaos. The fixed points are analyzed in Sec.~\ref{sec:statics}, and the corresponding magnetic structures are shown in Fig.~\ref{fig:static}ab. In the map for $\Lambda_1$ (Fig.~\ref{fig:Lambda1}) the parameter space domains which correspond to the fixed points A and B are marked in blue and are labeled ``FP$_{\textsc{a}}$'' and ``FP$_{\textsc{b}}$'', respectively. The white areas of the map for $\Lambda_1$ correspond to limit cycles and are labeled as ``LC''. Some examples of such limit cycles are studied in Sec.~\ref{sec:LC}, they are illustrated in Figs.~\ref{fig:dynamics}abc, see also insets I and II in Fig.~\ref{fig:transition}. Note that the line $\{j=0;b>1\}$ can be considered as a limit cycle of zero frequency. It is formed due to the continuous degeneration of the fixed point B with respect to the rotation of the vector $\vec{n}$ within $x-y$-plane. 

Red areas in the map for $\Lambda_1$ indicate chaotic dynamics; they are labeled as ``C''. Interestingly, in the vicinity of the spin-flop $b\lessapprox1$, the dynamics is mostly regular. This correlates with a high stability of the limit cycle \eqref{eq:lim-cycle} for the case $b=1$, see App.~\ref{app:Monodromy} for details. Also, as it follows from the shape of the LC-domain for lower fields, the current supports the stability of the limit cycle. Finally, the chaotic region is bounded for $b<1$. This is in contrast to the case $b>1$, where the current supports the emerging chaos. 

Note that at the boundaries between the periodic and chaotic regimes, there are quite narrow regions with signatures $\langle0,0,-,-\rangle$. They correspond to quasiperiodic dynamics on the surface of a 2-frequency tori in the phase space. Due to the narrowness, these quasiperiodic regions are not marked on the diagram in Fig.~\ref{fig:Lambda1}. Their investigation requires a significant increase in resolution of the LE maps and, therefore, high computing costs. Here, we restrict ourselves to only one example explored in Fig.~\ref{fig:transition}, where the transition to chaos goes through a relatively large region of quasiperiodic dynamics.

\begin{figure}
	\includegraphics[width=\columnwidth]{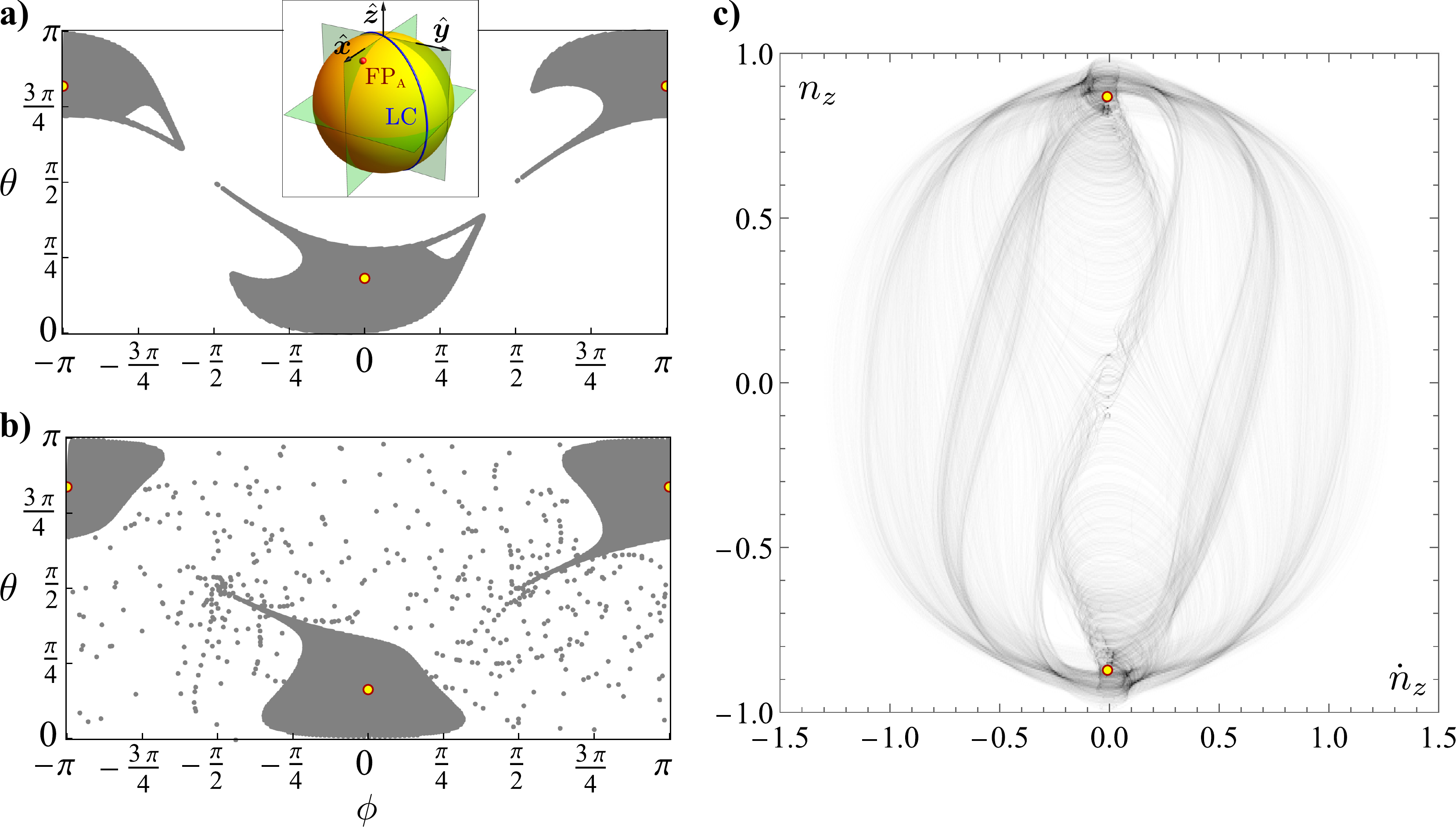}
	\caption{Basins of attraction of fixed point of type A (yellow spots) are shown by gray shadowing in panels a) and b). The 2D cross-section by the hyperplane $\{\dot{\phi}=0,\dot\theta=0\}$ of the 4D basins are presented. Panel a) corresponds to the case $(b,j)=(0.1,0.9)$ when the fixed point coexists with a limit cycle, which is shown in the inset for panel a). Panel b) corresponds to the case $(b,j)=(0.5,0.64)$ when the fixed point coexists with strange chaotic attractor, which is shown in panel c). The projection of the 4D strange attractor on the plane $(\dot{n}_z,n_z)$ is demonstrated. A natural measure of the strange attractor is shown by the level of gray tone. In all cases $\bar\alpha=0.1$.}\label{fig:basins}
\end{figure}

The maps of LEs were obtained for the phase trajectory with initial conditions 
\begin{equation}\label{eq:ini}
	\vec{n}(0)=\begin{cases}
		\hat{\vec{z}} & 0\le b\le1,\\
		\hat{\vec{x}} & b>1.
	\end{cases},\qquad \dot{\vec{n}}(0)=0.
\end{equation}
These initial conditions represent a possible experimental realization when the AFM sample is relaxed in the given magnetic field, and then the current is switched on step-by-step. If there is a single attractor in the phase space, then the choice of the trajectory (initial conditions) does not matter because, according to Oseledets theorem, all trajectories belonging to the same attractor have identical spectra of LEs. However, in our system, we found regimes when two attractors coexist, namely fixed point of type A coexists with limit cycle and with a strange attractor in regions ``LC/FP$_\textsc{a}$'', and ``C/FP$_\textsc{a}$'', respectively, see Fig. ~\ref{fig:Lambda1}. In these regimes, the long-term dynamics is sensitive to the initial conditions. We illustrate this by plotting the basins of attraction of the fixed point of type A for different regimes, see Fig.~\ref{fig:basins}. As it follows from panel a), one can reach the fixed point only by starting in the vicinity of this point (gray region). White points correspond to the attraction basin of the limit cycle. Note that the complete 4D attraction basins are path-connected, in contrast to the 2D-cross-sections presented here. 

The complicated structure of the attraction basin of the strange attractor (white area of panel b) allows us to conjecture that the boundary of the attraction basins has a fractal structure in 4D phase space \cite{Ott93}. The strange attractor is shown in Fig.~\ref{fig:basins}c) by means of its natural measure, which is the probability density to find the system in a given point of the phase space. Naturally, the highest probability density is in the vicinity of the fixed points.


In the map of $\Lambda_2$ one can distinguish a small region of hyperchaos where $\Lambda_2>0$, which is denoted as ``HC''. For example for the point $(b,j)=(1.18,0.54)$ we obtained $(\Lambda_1,\Lambda_2,\Lambda_3,\Lambda_4)=(0.017, 0.009, 0, -0.226)$ with an error smaller than $10^{-3}$. In order to better understand how chaos develops in this regime, we explore the behavior of the Lyapunov exponents along the line $b=1.18$ -- the vertical dashed line in Fig.~\ref{fig:Lambda1}. The development of chaos with increasing current is illustrated in Fig.~\ref{fig:transition} by analyzing the three largest LE. The smallest LE is always negative and not important for the analysis. For small currents $j<j_c\approx b^2-1$ all LEs are negative, which corresponds to the fixed point `B' illustrated in Fig.~\ref{fig:static}b). In the point `1', the two largest LE $\Lambda_1$ and $\Lambda_2$ are equal to each other and reach the value zero. As the current increases further, we obtain $\Lambda_1=0$ and $\Lambda_2<0$. This is the scenario where a soft-born, stable limit cycle (LC) emerges via an Andronov-Hopf bifurcation. When the current increases even further, the LC continuously increases its size and chances its form and position on the surface of the unit sphere, see inset I. In point `2', the LC experiences a jump-wise change of its form and position, see inset II. This is not a period-doubling bifurcation. After this, the two LEs $\Lambda_2$ and $\Lambda_3$ are equal to each other, reaching the value zero at point `3'. As the current keeps increasing, we have $\Lambda_1=\Lambda_2=0$ and $\Lambda_3<0$. This is the scenario where a soft-born, 2-frequency torus (T) emerges via a Neimark-Sacker bifurcation (NSB) \cite{Stankevich19,Vitolo11}. An example of the quasiperiodic dynamics on the torus surface is demonstrated in inset III by plotting the $\vec{n}$-vector trajectories obtained for the time interval $\Delta\tau=300$. In point `4' the torus changes its form and in point `5' the torus is destroyed, and chaos (C) appears, featuring a single positive LE. The latter process has a fine structure: the torus experiences synchronization, resulting in the creation of a LC, from which a subsequent torus emerges via a second NSB. In point `6', chaos is replaced by a periodic window, where a LC is present. In point `7' a 2-frequency torus bifurcates from the LC via NSB. With a small current increase, the torus is destroyed, and hyperchaos (HC) with two positive LEs develops. An example for hyperchaotic behavior is demonstrated on inset IV. Remarkably, a practically identical mechanism for the appearance of hyperchaos was recently reported for a Van der Pol oscillator with feedback loops \cite{Stankevich19}. A similar mechanism was also found for a system of two coupled R{\"o}{\ss}ler oscillators \cite{Kuznetsov15}. 

\begin{figure}
	\includegraphics[width=0.85\columnwidth]{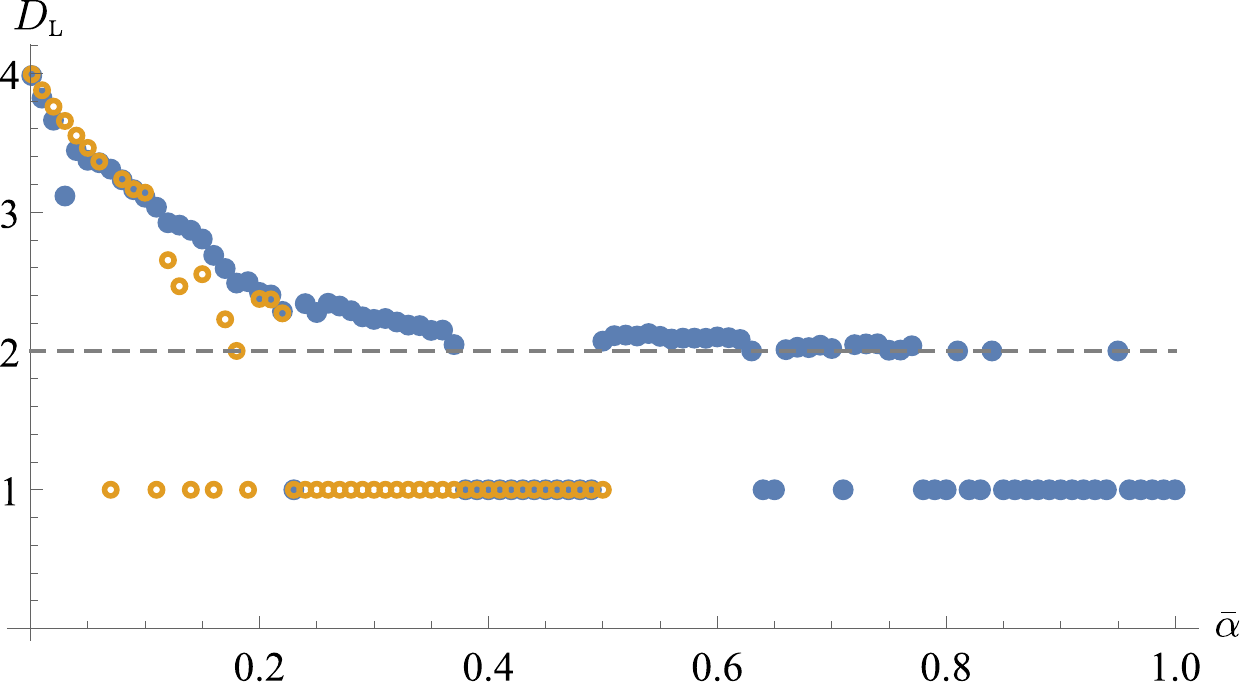}
	\caption{Dependence of the Lyapunov dimension of the chaotic attractor on the damping coefficient for the cases $(b,j)=(0.6,0.9)$ (open markers) and $(b,j)=(1.1,0.9)$ (closed markers).}\label{fig:DL}
\end{figure}
All the previous analysis was made for a constant damping $\bar{\alpha}=0.1$. In order to analyze how damping influences the chaotic dynamics, we consider the dependence of the Lyapunov dimension $D_{\textsc{l}}$ of the strange attractor on $\bar{\alpha}$ for the cases $b<1$ and $b>1$, see Fig.~\ref{fig:DL}. The Lyapunov dimension is used to approximate the informational dimension $D_1$ using the formula of Kaplan and Yorke:  
\begin{equation}\label{eq:KY}
	D_1\approx D_{\textsc{l}}=k+\frac{S_k}{|\Lambda_{k+1}|},
\end{equation}
where $k:$ $S_k\ge0$ and $S_{k+1}<0$.  In both cases, the behavior $D_{\textsc{l}}(\bar{\alpha})$ is the same if $\bar{\alpha}$ is small: for finite but vanishing $\bar{\alpha}$ one has $D_{\textsc{l}}\lessapprox d$ and both attractors demonstrate the same rate of decrease of $D_{\textsc{l}}$ with increasing $\bar{\alpha}$. However, for larger damping the behavior is different. In the case $b<1$, chaos is suppressed already for $\bar{\alpha}\approx0.2$ when the chaotic attractor tranforms into a limit cycle. In the case $b>1$, chaotic dynamics -- intercepted by periodic windows -- exists for much larger dampings, up to $\bar{\alpha}\approx0.8$. In the latter case, the strange attractor becomes almost flat. 



\begin{figure}
	\includegraphics[width=\columnwidth]{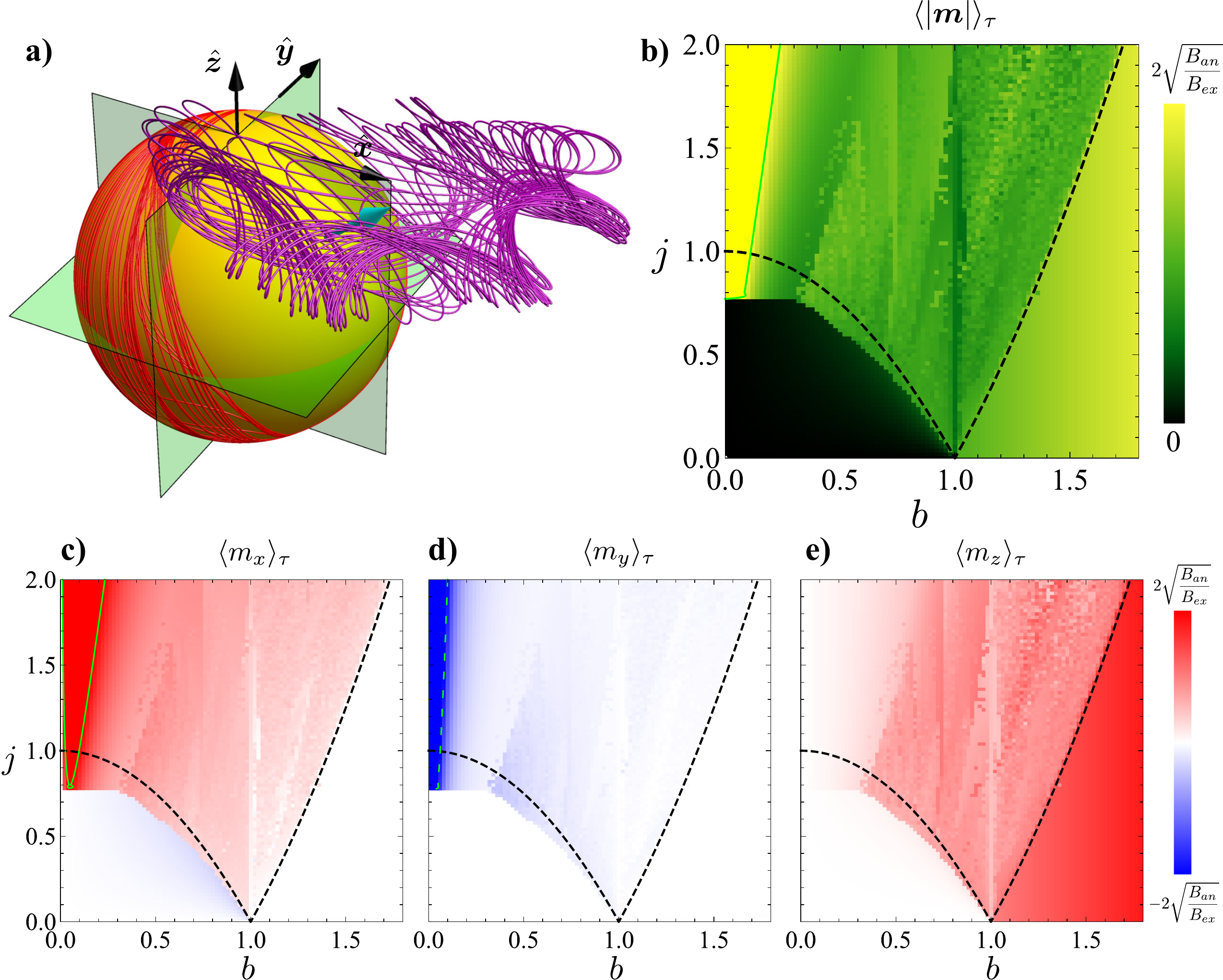}
	\caption{Interpretation of the dynamical regimes in terms of the magnetization. a) -- examples of chaotic trajectories made by $\vec{n}$- (red) and $\vec{m}$-vectors (purple) for the case $b=0.5$, $j=1$, $\alpha=0.1$. The cyan arrow shows the averaged magnetization. Panels b), c), d) and e) show the  absolute value and the three Cartesian components of the averaged magnetization, respectively. Solid and dashed green lines show the clipping values $2\sqrt{B_{an}/B_{ex}}$ and $-2\sqrt{B_{an}/B_{ex}}$, respectively. The maximum value $\langle |\vec{m}|\rangle_\tau=\bar{\alpha}^{-1}\sqrt{B_{an}/B_{ex}}$ is reached in point $(b,j)=(0,2)$, compare with \eqref{eq:m-lim-cycle} and  \eqref{eq:lim-cycle}.}\label{fig:mag}
\end{figure}

\emph{Chaos characterization by analyzing the magnetization.} The observation of the dynamics of the N{\'e}el vector is a challenging experimental problem. On the contrary, the magnetization dynamics can be easily observed by detecting the generated ac magnetic field. For this reason, it is instructive to explore the relation between the given regime of the system dynamics and the behavior of the magnetization vector \eqref{eq:m}. In Fig.~\ref{fig:dynamics}a-c we demonstrated how the Fourier spectra of a magnetization component can be used to distinguish regular and chaotic dynamics. Here, we analyzed the behavior of the magnetization vector for a wide range of parameters space, see Fig.~\ref{fig:mag}. The magnetizations $\vec{m}_{\textsc{a}}$ and $\vec{m}_{\textsc{a}}$ of the static states (A) and (B) are proportional to the applied magnetic field (see Sec.~\ref{sec:statics}), wherein $|\vec{m}_{\textsc{b}}|$ is noticeably larger and exactly parallel to the applied magnetic field. The time-averaged magnetization generated by the low-field periodic dynamics is determined by the first summand in \eqref{eq:m-lim-cycle}. It is oriented within the plane perpendicular to the field. In chaotic regimes, the magntization dynamics is rather complicated, see Fig.~\ref{fig:mag}a). Remarkably, the time-averaged magnetization $\langle \vec{m}\rangle_\tau$ is determined only by the control parameters $b$, $j$, $\bar{\alpha}$, and it is a property of the attractor realized for the given dynamics, see Fig.~\ref{fig:mag}b)-e). The spatial octant, in which vector $\langle \vec{m}\rangle_\tau$ lies is determined by vectors $\vec{b}$ and $\vec{p}$, namely $\text{sign}(\langle m_x\rangle_\tau)=\text{sign}(bp)$, $\text{sign}(\langle m_y\rangle_\tau)=\text{sign}(p)$, and $\text{sign}(\langle m_z\rangle_\tau)=\text{sign}(b)$. Here, vectors $\vec{b}$ and $\vec{p}$ are applied along axes $\hat{\vec{z}}$ and $\hat{\vec{y}}$, respectively.

\section{Conclusions}
The simultaneous action of magnetic field and spin-torque on AFM nano-oscillator can result in complicated nonlinear dynamics of the AFM order parameter. Depending on the control parameters ($b$, $j$), the system demonstrates different types of nonlinear behavior: regular dynamics along limit circles, quasiperiodic dynamics along 2-frequency tori, as well as chaotic dynamics. The latter can be of interest for neuromorphic computing relying on stochastic elements. We show that the threshold current of the appearance of chaos is especially low in the vicinity of the spin-flop transition.  

\section{Acknowledgments}
B.W. and V.K. thank Markus Garst for fruitful discussions and acknowledge support from DFG Project-ID 270344603, and 324327023. O.G. acknowledges support from the Alexander von Humboldt Foundation, the ERC Synergy Grant SC2 (No. 610115), and the Deutsche Forschungsgemeinschaft (DFG, German Research Foundation) - TRR 173 – 268565370 (project A11). In part, this work was supported by the Program of Fundamental Research of the Department of Physics and Astronomy of the National Academy of Sciences of Ukraine (Project No. 0116U003192).

\appendix
\section{Structure of hamiltonian and simplified equations of motion.}\label{app:eqs}
This Appendix aims to introduce the number of material parameters e.g. exchange $B_{ex}$ and anisotropy $B_{an}$ fields, and also to provide the derivation of the equations of motion in exchange approximation $B_{ex}\gg B_{an}$. The latter derivation was previously made in Ref.~\onlinecite{Gomonay10}, and we provide it here for the sake of text coherence. For the derivation without spin torques, see also Refs.~\onlinecite{Baryakhtar79,Ivanov95e,Turov01en}.

We assume that the energy density of the system can be presented in the form $E=E_{ex}+E_{an}+E_{zee}$. Here, the first summand represents the exchange contribution with density \cite{Kaganov58,Akhiezer68}
\begin{equation}\label{eq:Eex}
	\begin{split}
	&\mathcal{E}_{ex}=\frac{B_{ex}M_s}{2}\left(\vec{\mu}_1\cdot\vec{\mu}_2\right)\\
	&+\sum\limits_{i=x,y,z}\left\{\frac{A'}{2}\left[(\partial_i\vec\mu_1)^2+(\partial_i\vec\mu_2)^2\right]+A''\left(\partial_i\vec{\mu}_1\cdot\partial_i\vec{\mu}_2\right)\right\},
	\end{split}
\end{equation} 
which includes uniform exchange between the sublattices ($B_{ex}$) and isotropic nonuniform intra- ($A'$) and inter- ($A''$) lattices exchange interactions. In terms of the vectors $\vec{n}$ and $\vec{m}$ the exchange energy density \eqref{eq:Eex} has the form
\begin{equation}\label{eq:Eex-mn}
	\mathcal{E}_{ex}=B_{ex}M_s\left(\vec{m}^2-\frac{1}{2}\right)+\sum\limits_{i=x,y,z}\left\{A(\partial_i\vec{n})^2+\tilde{A}(\partial_i\vec{m})^2\right\},
\end{equation}
where $A=A'-A''$ and $\tilde{A}=A'+A''$.

The energy of uniaxial anisotropy $E_{an}$ has the density \cite{Kaganov58,Akhiezer68}
\begin{equation}\label{eq:Ean}
	\begin{split}
	\mathcal{E}_{an}=&-\frac{K'}{2}\left[(\vec{\mu}_1\cdot\hat{\vec{z}})^2+(\vec{\mu}_2\cdot\hat{\vec{z}})^2\right]-K''(\vec{\mu}_1\cdot\hat{\vec{z}})(\vec{\mu}_2\cdot\hat{\vec{z}})\\
	=&-M_s\left[B_{an}(\vec{n}\cdot\hat{\vec{z}})^2+\tilde{B}_{an}(\vec{m}\cdot\hat{\vec{z}})^2\right],
	\end{split}
\end{equation}
where we assumed that the anisotropy easy-axis is oriented along $\hat{\vec{z}}$, and also introduced the anisotropy fields $B_{an}=(K'-K'')/M_s$ and $\tilde{B}_{an}=(K'+K'')/M_s$. 

Finally, the interaction with an external magnetic field $\vec{B}$ is
\begin{equation}\label{eq:Ezee}
	\mathcal{E}_{zee}=-M_s\vec{B}\cdot(\vec{\mu}_1+\vec{\mu}_2)=-2M_s\vec{B}\cdot\vec{m}.
\end{equation}

In terms of the vectors $\vec{n}$ and $\vec{m}$ the set of Landau-Lifshitz equations takes the following form
\begin{subequations}\label{eq:LL-mn}
	\begin{align}
\label{eq:LL-n}\dot{\vec{n}}&=\frac{\gamma}{2M_s}\left[\vec{m}\times\frac{\delta E}{\delta\vec{n}}+\vec{n}\times\frac{\delta E}{\delta\vec{m}}\right]\\
&+\alpha\left[\vec{m}\times\dot{\vec{n}}+\vec{n}\times\dot{\vec{m}}\right]	-\sigma\left[\vec{m}(\vec{p}\cdot\vec{n})+\vec{n}(\vec{p}\cdot\vec{m})\right], \nonumber \\
\label{eq:LL-m}\dot{\vec{m}}&=\frac{\gamma}{2M_s}\left[\vec{m}\times\frac{\delta E}{\delta\vec{m}}+\vec{n}\times\frac{\delta E}{\delta\vec{n}}\right]\\
&+\alpha\left[\vec{m}\times\dot{\vec{m}}+\vec{n}\times\dot{\vec{n}}\right]+\sigma\left[\vec{p}-\vec{m}(\vec{p}\cdot\vec{m})-\vec{n}(\vec{p}\cdot\vec{n})\right].  \nonumber
\end{align}
\end{subequations}
Based on the form of the energy functional $E$, we conclude that in the static case $|\vec{m}|\sim B/B_{ex}\ll1$ for a large exchange field $B_{ex}\gg B,\tilde{B}_{an}$. In this limit, we neglect in Eq.~\eqref{eq:LL-n} all terms proportional to $\vec{m}$ except the leading linear term $\propto B_{ex}\vec{m}$. From this we find that $\dot{\vec{n}}\approx\gamma\vec{n}\times\left[\vec{m}B_{ex}-\vec{B}\right]$, where it was also assumed that $B_{ex}\gg B_{an}$. Using this in the considered limit $|\vec{n}|\approx1$, 
we solve the latter equation with respect to $\vec{m}$:
\begin{equation}\label{eq:m-appr}
	\vec{m}\approx\frac{1}{B_{ex}}\left(\gamma^{-1}\dot{\vec{n}}+\vec{n}\times\vec{B}\right)\times\vec{n}
\end{equation}
Now we take the time derivative of Eq.~\eqref{eq:m-appr}, substitute $\dot{\vec{m}}$ into \eqref{eq:LL-m} and neglect all terms with higher order than $\mathcal{O}(B_{ex}^{-1})$. The resulting equation reads
\begin{equation}\label{eq:n}
	\begin{split}
	\Bigl[&\ddot{\vec{n}}+\gamma(\vec{n}\times\dot{\vec{B}})+2\gamma\left(\dot{\vec{n}}\times\vec{B}\right)+\gamma^2\vec{B}(\vec{n}\cdot\vec{B})\\
	&+\frac{\gamma^2B_{ex}}{2M_s}\frac{\delta E}{\delta\vec{n}}+\alpha\gamma B_{ex}\dot{\vec{n}}-\sigma\gamma B_{ex}(\vec{n}\times\vec{p})\Bigr]\times\vec{n}=0
	\end{split}
\end{equation}
For the special case of uniform magnetization $\vec{n}=\vec{n}(t)$, magnetic field applied along the anisotropy axis $\vec{B}=B\hat{\vec{z}}$, and $\vec{p}=-\hat{\vec{y}}$, Eq.~\eqref{eq:n} is transformed to \eqref{eq:n-dyn} if we utilize the dimensionless units explained in the main text. In terms of dimensionless units, Eq.~\eqref{eq:m-appr} is transformed to \eqref{eq:m}.

\section{Stability analysis of static fixed points.}\label{app:stability}

\begin{figure*}
	\centering
	\includegraphics[width=\linewidth]{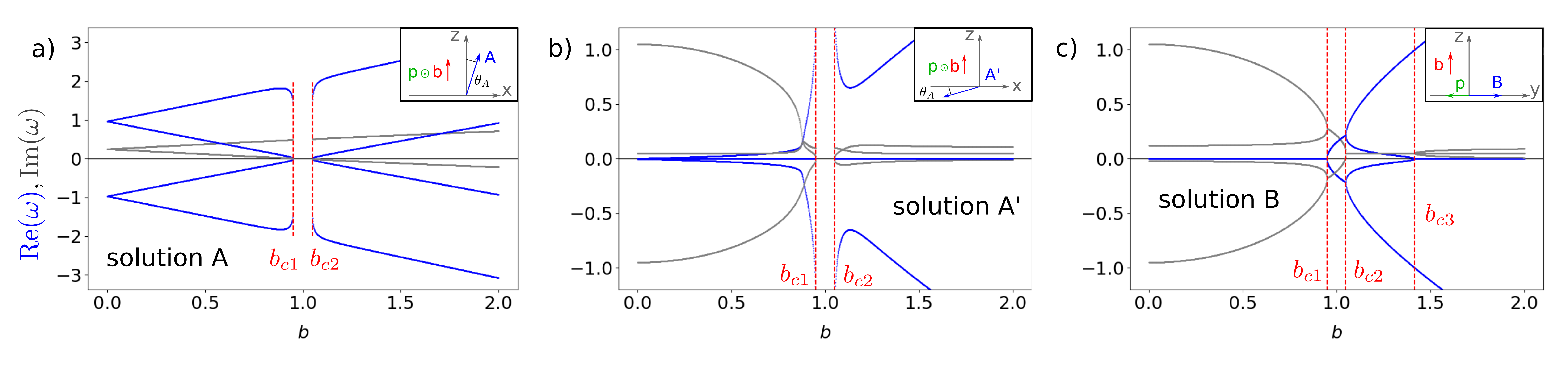}   
	\caption{Evolution of the eigenvalue spectrum of solutions (A), (A$^\prime$) and (B) with the magnetic field $b$ for the current $j = 0.1$: $\mathrm{Re}(\omega)$ is plotted in blue, $\mathrm{Im}(\omega)$ is plotted in grey. In subfigure a) for solution (A) $\bar{\alpha} = 0.5$ was chosen to demonstrate the splitting in the imaginary part of $\omega$. While for $b < 1.0$ both branches are positive, for $b > 1.0$ one branch becomes negative, indicating the instability of solution (A) above the second critical magnetic field. In subfigure b) for solution (A$^\prime$) and $\bar{\alpha} = 0.1$ there is always one negative branch of the imaginary part, i.e. it is never stable. And in subfigure c) for solution (B) and $\bar{\alpha} = 0.1$ the imaginary part becomes negative for $b \lessapprox b_{c2}$, slightly below the second critical magnetic field.}
	\label{fig:eigenvaluespectrum}
\end{figure*}

The equations of motion \eqref{eq:angles} can be rewritten as a first-order differential problem $\dot{\vec{x}} = \vec{f}(\vec{x})$ with $\vec{x} = (\theta,\dot{\theta},\phi,\dot{\phi})^{\textsc{t}}$.
%
%
There are six different fixed points $\vec{x}^*:$ $\vec{f}(\vec{x}^*) = \vec{0}$,
which correspond to three distinct physical states, since $\vec{n}\to-\vec{n}$. The three static, uniform solutions are visualized in Fig.~\ref{fig:eigenvaluespectrum}: for finite field $b$ and current $j$ solutions (A) and (A$^\prime$) are tilted off the $z$- and $x$-axis, respectively, by an angle $\theta_{\textsc{A}}$. At the critical current $j_c$, both fixed points collide and annihilate in a saddle-node-bifurcation at $\theta_{\textsc{A}} = \pi/4$. Solution (B) persists for all parameters and is always oriented parallel/anti-parallel to the direction of spin polarization $\vec{p}$. 

The stability of a fixed point $\vec{x}^*$ can be determined by solving the linearized system  $\dot{\tilde{\vec{x}}} = \hat{L} \tilde{\vec{x}}$ with $\hat{L} = \frac{\partial \vec{f}}{\partial \vec{x}} \Big|_{\vec{x}=\vec{x}^*}$ being the constant matrix and $\tilde{\vec{x}}=\vec{x}-\vec{x}^*$ are small perturbations. Solutions have the form $\tilde{\vec{x}}(\tau) = \tilde{\vec{x}}_0 e^{i \omega \tau}$, where $\omega=-i\lambda$ and $\lambda$ is an eigenvalue of $\hat{L}$. 
%
%
%

For solution A the eigenvalues are cumbersome expressions, but for solution B, they can be determined more easily from
\begin{equation}
	\hat{L}_{\textsc{b}} = 
	\begin{pmatrix}
		0 & 1 & 0 & 0 \\
		1-b^2 & -\bar{\alpha}  & -\frac{j}{2} & 0 \\
		0 & 0 & 0 & 1 \\
		\frac{j}{2} & 0 & 0 & -\bar{\alpha} \\
	\end{pmatrix}
\end{equation}
leading to
\begin{equation}
	\omega = \frac{i\bar{\alpha}}{2} \pm \sqrt{\left(\omega_{\textsc{h},\textsc{l}}^{\textsc{b}}\right)^2- \bar{\alpha}^2/4},
\end{equation}
where the eigenfrequencies $\omega_{\textsc{h}}^{\textsc{b}}$ and $\omega_{\textsc{l}}^{\textsc{b}}$ are defined in Sec.~\ref{sec:statics}.
The evolution of $\omega$ with $b$ for finite damping $\bar{\alpha}$ is shown in Fig. \ref{fig:eigenvaluespectrum}.

For solution (A) $\bar{\alpha} = 0.5$ was chosen in order to feature the splitting in the imaginary part. For $b < b_{c1}$, the two branches $\text{Im}(\omega)$ are both positive and they separate linearly with $b$. For $b > b_{c2}$, the splitting is still linear in $b$, but now one of the branches is negative, indicating the instability of solution (A). This branch is negative only for finite damping $\bar{\alpha}$, a phenomenon which is referred to as "dynamic instability" and which is known also e.g. from structural mechanics \cite{Mascolo19}. 

For solution (A$^\prime$) there is always a negative branch of the imaginary part of $\omega$, thus solution (A$^\prime$) is never stable.

For $b < b_{c2}$ there is always at least one negative branch of $\text{Im}(\omega)$ of solution (B) and thus it is unstable. At $b_{c1}$ pairs of eigenvalues collide on the imaginary axis in a reversible Hopf-bifurcation. 
This is analogous to the upward Ziegler double pendulum, which displays also a two-by-two reversible Hopf-bifurcation, depending on the follower load \cite{Ziegler52,Mascolo19}. In the field range $b_{c1} < b < b_{c2}$ solution (B) experiences the flutter instability.
Solution (B) gains stability at $b_{c2}(\bar{\alpha})$. For $b > b_{c2}$ there is at first a single, positive branch of $\text{Im}(\omega)$, which is then splitting at $b_{c3} = \sqrt{1 + \frac{\bar{\alpha}^4 + 4 j^2}{4 \bar{\alpha}^2}}$ into three branches. Here, $b_{c3}$ indicates the threshold, where the in-plane, linear oscillations become overdamped, as the corresponding branch of the real part becomes zero at $b_{c3}$.

\section{Stability analysis of static fixed points.}\label{app:stability}

The equations of motion \eqref{eq:angles} can be rewritten as a first-order differential problem $\dot{\vec{x}} = \vec{f}(\vec{x})$ with $\vec{x} = (\theta,\dot{\theta},\phi,\dot{\phi})^{\textsc{t}}$.
%
%
There are six different fixed points $\vec{x}^*$: $\vec{f}(\vec{x}^*) = \vec{0}$, which correspond to three distinct physical states, since $\vec{n}\to-\vec{n}$ is a symmetry. The three static, uniform solutions are visualized in Fig.~\ref{fig:eigenvaluespectrum}: for finite field $b$ and current $j$ solutions (A) and (A$^\prime$) are tilted off the $z$- and $x$-axis, respectively, by an angle $\theta_{\textsc{A}}$. At the critical current $j_c$, both fixed points collide and annihilate in a saddle-node-bifurcation at $\theta_{\textsc{A}} = \pi/4$. The solution (B) persists for all parameters and is always oriented parallel/anti-parallel to the direction of spin polarization $\vec{p}$. 

The stability of a fixed point $\vec{x}^*$ can be determined by solving the linearized system  $\dot{\tilde{\vec{x}}} = \hat{L} \tilde{\vec{x}}$ with $\hat{L} = \frac{\partial \vec{f}}{\partial \vec{x}} \Big|_{\vec{x}=\vec{x}^*}$ being the constant matrix and $\tilde{\vec{x}}=\vec{x}-\vec{x}^*$ are small perturbations. The solutions have the form $\tilde{\vec{x}}(\tau) = \tilde{\vec{x}}_0 e^{i \omega \tau}$, where $\omega=-i\lambda$ and $\lambda$ is an eigenvalue of $\hat{L}$. 
%
%
%

For solution A the eigenvalues are cumbersome expressions, but for solution B, they can be determined more easily from
\begin{equation}
	\hat{L}_{\textsc{b}} = 
	\begin{pmatrix}
		0 & 1 & 0 & 0 \\
		1-b^2 & -\bar{\alpha}  & -\frac{j}{2} & 0 \\
		0 & 0 & 0 & 1 \\
		\frac{j}{2} & 0 & 0 & -\bar{\alpha} \\
	\end{pmatrix}
\end{equation}
leading to
\begin{equation}
	\omega = \frac{i\bar{\alpha}}{2} \pm \sqrt{\left(\omega_{\textsc{h},\textsc{l}}^{\textsc{b}}\right)^2- \bar{\alpha}^2/4},
\end{equation}
where the eigenfrequencies $\omega_{\textsc{h}}^{\textsc{b}}$ and $\omega_{\textsc{l}}^{\textsc{b}}$ are defined in Sec.~\ref{sec:statics}.
The evolution of $\omega$ with $b$ for finite damping $\bar{\alpha}$ is shown in Fig. \ref{fig:eigenvaluespectrum}.

For solution (A) $\bar{\alpha} = 0.5$ was chosen in order to feature the splitting in the imaginary part. For $b < b_{c1}$, the two branches $\text{Im}(\omega)$ are both positive and they separate linearly with $b$. For $b > b_{c2}$, the splitting is still linear in $b$, but now one of the branches is negative, indicating the instability of solution (A). This branch is negative only for finite damping $\bar{\alpha}$, a phenomenon which is referred to as "dynamic instability" and which is known also e.g. from structural mechanics \cite{Mascolo19}. 

For solution (A$^\prime$) there is always a negative branch of the imaginary part of $\omega$, thus, solution (A$^\prime$) is never stable.

For $b < b_{c2}$ there is always at least one negative branch of $\text{Im}(\omega)$ of solution (B) and thus it is unstable. At $b_{c1}$ pairs of eigenvalues collide on the imaginary axis in a reversible Hopf-bifurcation. 
This is analogous to the upward Ziegler double pendulum, which displays also a two-by-two reversible Hopf-bifurcation, depending on the follower load \cite{Ziegler52,Mascolo19}. In the field range $b_{c1} < b < b_{c2}$ solution (B) experiences a flutter instability.
It gains stability at $b_{c2}(\bar{\alpha})$. For $b > b_{c2}$ there is at first a single, positive branch of $\text{Im}(\omega)$, which is then splitting at $b_{c3} = \sqrt{1 + \frac{\bar{\alpha}^4 + 4 j^2}{4 \bar{\alpha}^2}}$ into three branches. Here, $b_{c3}$ indicates the threshold, where the in-plane, linear oscillations become overdamped, as the corresponding branch of the real part becomes zero at $b_{c3}$.

\section{Stability analysis of the limit cycle for $b=1$.}\label{app:Monodromy}

\begin{figure*}
	\centering
	\includegraphics[width=\linewidth]{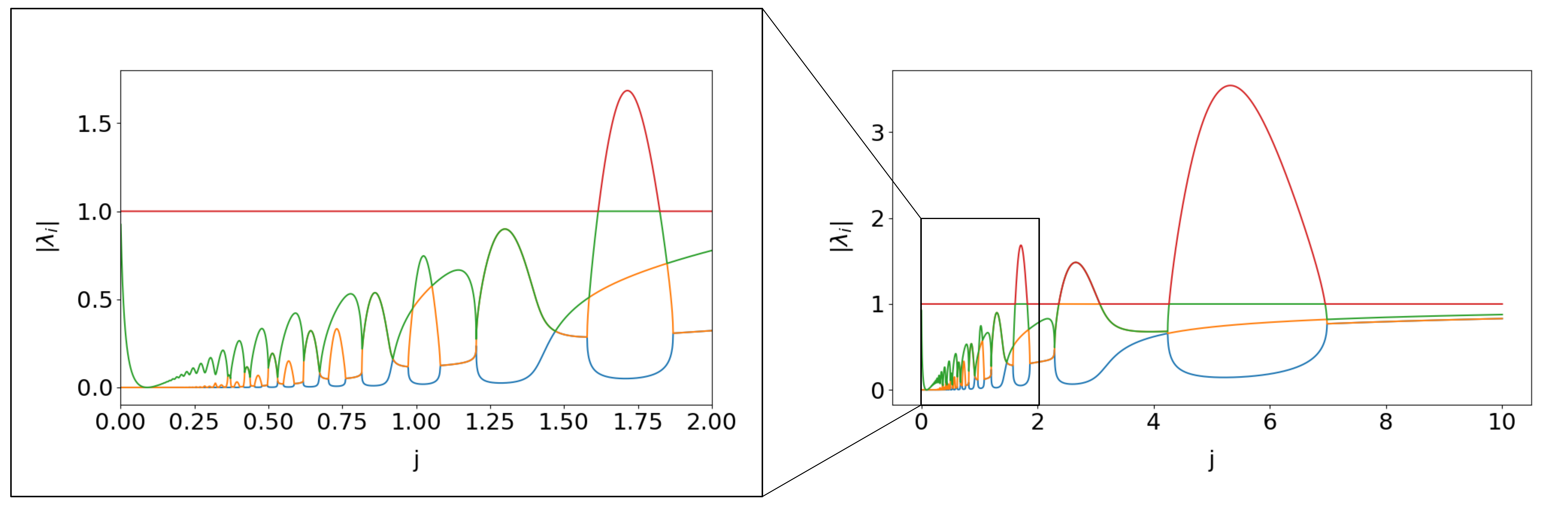}	
	\caption{Evolution of the Floquet multipliers (absolute value $|\lambda_i|$ of the monodromy matrix) for the limit cycle at $b = 1.0$ and $\alpha = 0.1$. The limit cycle is remarkable stable with just one island of instability $1.615 < j < 1.824$, where the greatest Floquet multiplier is larger than one. There are only two more areas of instability for larger current: $2.366 < j < 3.073$ and $4.265 < j < 6.953$.}
	\label{fig:monodromyspectrum}
\end{figure*}

The stability of the limit cycle at $b = 1$ can be analyzed within the theory of Floquet multipliers, i.e. the absolute value of the eigenvalues $|\lambda_i|$ of the limit cycle's monodromy matrix $\hat{M}$. If one Floquet multiplier is greater than one, i.e. $|\lambda_i| > 1$, the limit cycle becomes unstable \cite{Argyris15}.

The limit cycle was derived in section IV and for $b = 1$ this is an exact solution of the equations of motion with
\begin{equation}\label{eq:exact-lim-cycle}
	\phi_0=\arctan(2/\bar{\alpha}),\qquad \Omega_0=\frac{j}{2\sqrt{\bar{\alpha}^2+4}}.
\end{equation} 
Since this limit cycle passes through the poles of the unit sphere, it is necessary to work in a rotated frame of reference, in order to avoid singularities, where $\tilde{\vec{n}} = \sin\tilde{\theta} \left( \sin\tilde{\phi}\, \hat{\vec{x}} + \cos\tilde{\phi}\,\hat{\vec{z}} \right) + \cos\tilde{\theta} \hat{\vec{y}}$ is the N{\'e}el vector in the rotated frame of reference. The exact solution for $\tilde{\theta}(\tau)$ and $\tilde{\phi}(\tau)$ can be obtained from the relation $\tilde{\vec{n}}=\vec{n}$, where $\vec{n}=\sin(\Omega_0\tau)(\cos\phi_0\hat{\vec{x}}+\sin\phi_0\hat{\vec{y}})+\cos(\Omega_0\tau)\hat{\vec{z}}$.

The monodromy matrix $\hat{M} = \hat{\Phi}(T)$ corresponds to the fundamental matrix $\hat{\Phi}$ of the linearized dynamical system $\dot{\tilde{\vec{x}}} = \hat{L} \tilde{\vec{x}}$, evaluated at the orbit period $T = 2 \pi / \Omega_0$ of the limit cycle. In order to determine it, the matrix $\hat{L} = \frac{\partial \vec{f}}{\partial \vec{x}} \Big|_{\vec{x}=\vec{x}_{\mathrm{cycle}}}$ is evaluated at the exactly know solution of the limit cycle, in the rotated frame of reference. Next, the linearized system is solved numerically, starting from $\hat{\Phi}(0) = \hat{1}$, for one orbit period $T$. 

The absolute value of the eigenvalues of the resulting monodromy matrix $\hat{M}$ are the Floquet multipliers; their evolution is shown in Fig. \ref{fig:monodromyspectrum} depending on the current $j$.

The limit cycle is remarkable stable, having only three instability islands, where the greatest Floquet multiplier is larger than one, see Fig.~\ref{fig:monodromyspectrum}. The first instability island $1.615 < j < 1.824$ is consistent with a small island of chaos, where the greatest Lyapunov exponent is greater than zero in Fig. \ref{fig:Lambda1}. There are only two more areas of instability for larger current: $2.366 < j < 3.073$ and $4.265 < j < 6.953$. 

Remarkably, in the case $j=0$, we have $|\lambda_1|=|\lambda_2|=1$. This means that $\Lambda_1=\Lambda_2=0$ and, thus, it corresponds to quasiperiodic dynamics on a torus in phase space.

The presence of this stable limit cycle explains why chaotic dynamics is almost absent along the line $b=1$ in the Lyapunov spectra, see Fig.~\ref{fig:Lambda1}.


\begin{thebibliography}{41}%
	\makeatletter
	\providecommand \@ifxundefined [1]{%
		\@ifx{#1\undefined}
	}%
	\providecommand \@ifnum [1]{%
		\ifnum #1\expandafter \@firstoftwo
		\else \expandafter \@secondoftwo
		\fi
	}%
	\providecommand \@ifx [1]{%
		\ifx #1\expandafter \@firstoftwo
		\else \expandafter \@secondoftwo
		\fi
	}%
	\providecommand \natexlab [1]{#1}%
	\providecommand \enquote  [1]{``#1''}%
	\providecommand \bibnamefont  [1]{#1}%
	\providecommand \bibfnamefont [1]{#1}%
	\providecommand \citenamefont [1]{#1}%
	\providecommand \href@noop [0]{\@secondoftwo}%
	\providecommand \href [0]{\begingroup \@sanitize@url \@href}%
	\providecommand \@href[1]{\@@startlink{#1}\@@href}%
	\providecommand \@@href[1]{\endgroup#1\@@endlink}%
	\providecommand \@sanitize@url [0]{\catcode `\\12\catcode `\$12\catcode
		`\&12\catcode `\#12\catcode `\^12\catcode `\_12\catcode `\%12\relax}%
	\providecommand \@@startlink[1]{}%
	\providecommand \@@endlink[0]{}%
	\providecommand \url  [0]{\begingroup\@sanitize@url \@url }%
	\providecommand \@url [1]{\endgroup\@href {#1}{\urlprefix }}%
	\providecommand \urlprefix  [0]{URL }%
	\providecommand \Eprint [0]{\href }%
	\providecommand \doibase [0]{http://dx.doi.org/}%
	\providecommand \selectlanguage [0]{\@gobble}%
	\providecommand \bibinfo  [0]{\@secondoftwo}%
	\providecommand \bibfield  [0]{\@secondoftwo}%
	\providecommand \translation [1]{[#1]}%
	\providecommand \BibitemOpen [0]{}%
	\providecommand \bibitemStop [0]{}%
	\providecommand \bibitemNoStop [0]{.\EOS\space}%
	\providecommand \EOS [0]{\spacefactor3000\relax}%
	\providecommand \BibitemShut  [1]{\csname bibitem#1\endcsname}%
	\let\auto@bib@innerbib\@empty
	\bibitem [{\citenamefont {Markovi{\'{c}}}\ \emph {et~al.}(2020)\citenamefont
		{Markovi{\'{c}}}, \citenamefont {Mizrahi}, \citenamefont {Querlioz},\ and\
		\citenamefont {Grollier}}]{Markovic20}%
	\BibitemOpen
	\bibfield  {author} {\bibinfo {author} {\bibfnamefont {Danijela}\
			\bibnamefont {Markovi{\'{c}}}}, \bibinfo {author} {\bibfnamefont {Alice}\
			\bibnamefont {Mizrahi}}, \bibinfo {author} {\bibfnamefont {Damien}\
			\bibnamefont {Querlioz}}, \ and\ \bibinfo {author} {\bibfnamefont {Julie}\
			\bibnamefont {Grollier}},\ }\bibfield  {title} {\enquote {\bibinfo {title}
			{Physics for neuromorphic computing},}\ }\href {\doibase
		10.1038/s42254-020-0208-2} {\bibfield  {journal} {\bibinfo  {journal} {Nature
				Reviews Physics}\ }\textbf {\bibinfo {volume} {2}},\ \bibinfo {pages}
		{499--510} (\bibinfo {year} {2020})}\BibitemShut {NoStop}%
	\bibitem [{\citenamefont {Roy}\ \emph {et~al.}(2019)\citenamefont {Roy},
		\citenamefont {Jaiswal},\ and\ \citenamefont {Panda}}]{Roy19}%
	\BibitemOpen
	\bibfield  {author} {\bibinfo {author} {\bibfnamefont {Kaushik}\ \bibnamefont
			{Roy}}, \bibinfo {author} {\bibfnamefont {Akhilesh}\ \bibnamefont {Jaiswal}},
		\ and\ \bibinfo {author} {\bibfnamefont {Priyadarshini}\ \bibnamefont
			{Panda}},\ }\bibfield  {title} {\enquote {\bibinfo {title} {Towards
				spike-based machine intelligence with neuromorphic computing},}\ }\href
	{\doibase 10.1038/s41586-019-1677-2} {\bibfield  {journal} {\bibinfo
			{journal} {Nature}\ }\textbf {\bibinfo {volume} {575}},\ \bibinfo {pages}
		{607--617} (\bibinfo {year} {2019})}\BibitemShut {NoStop}%
	\bibitem [{\citenamefont {Segall}\ \emph {et~al.}(2017)\citenamefont {Segall},
		\citenamefont {LeGro}, \citenamefont {Kaplan}, \citenamefont {Svitelskiy},
		\citenamefont {Khadka}, \citenamefont {Crotty},\ and\ \citenamefont
		{Schult}}]{Segall17}%
	\BibitemOpen
	\bibfield  {author} {\bibinfo {author} {\bibfnamefont {K.}~\bibnamefont
			{Segall}}, \bibinfo {author} {\bibfnamefont {M.}~\bibnamefont {LeGro}},
		\bibinfo {author} {\bibfnamefont {S.}~\bibnamefont {Kaplan}}, \bibinfo
		{author} {\bibfnamefont {O.}~\bibnamefont {Svitelskiy}}, \bibinfo {author}
		{\bibfnamefont {S.}~\bibnamefont {Khadka}}, \bibinfo {author} {\bibfnamefont
			{P.}~\bibnamefont {Crotty}}, \ and\ \bibinfo {author} {\bibfnamefont
			{D.}~\bibnamefont {Schult}},\ }\bibfield  {title} {\enquote {\bibinfo {title}
			{Synchronization dynamics on the picosecond time scale in coupled josephson
				junction neurons},}\ }\href {\doibase 10.1103/physreve.95.032220} {\bibfield
		{journal} {\bibinfo  {journal} {Physical Review E}\ }\textbf {\bibinfo
			{volume} {95}},\ \bibinfo {pages} {032220} (\bibinfo {year}
		{2017})}\BibitemShut {NoStop}%
	\bibitem [{\citenamefont {Torrejon}\ \emph {et~al.}(2017)\citenamefont
		{Torrejon}, \citenamefont {Riou}, \citenamefont {Araujo}, \citenamefont
		{Tsunegi}, \citenamefont {Khalsa}, \citenamefont {Querlioz}, \citenamefont
		{Bortolotti}, \citenamefont {Cros}, \citenamefont {Yakushiji}, \citenamefont
		{Fukushima}, \citenamefont {Kubota}, \citenamefont {Yuasa}, \citenamefont
		{Stiles},\ and\ \citenamefont {Grollier}}]{Torrejon17}%
	\BibitemOpen
	\bibfield  {author} {\bibinfo {author} {\bibfnamefont {Jacob}\ \bibnamefont
			{Torrejon}}, \bibinfo {author} {\bibfnamefont {Mathieu}\ \bibnamefont
			{Riou}}, \bibinfo {author} {\bibfnamefont {Flavio~Abreu}\ \bibnamefont
			{Araujo}}, \bibinfo {author} {\bibfnamefont {Sumito}\ \bibnamefont
			{Tsunegi}}, \bibinfo {author} {\bibfnamefont {Guru}\ \bibnamefont {Khalsa}},
		\bibinfo {author} {\bibfnamefont {Damien}\ \bibnamefont {Querlioz}}, \bibinfo
		{author} {\bibfnamefont {Paolo}\ \bibnamefont {Bortolotti}}, \bibinfo
		{author} {\bibfnamefont {Vincent}\ \bibnamefont {Cros}}, \bibinfo {author}
		{\bibfnamefont {Kay}\ \bibnamefont {Yakushiji}}, \bibinfo {author}
		{\bibfnamefont {Akio}\ \bibnamefont {Fukushima}}, \bibinfo {author}
		{\bibfnamefont {Hitoshi}\ \bibnamefont {Kubota}}, \bibinfo {author}
		{\bibfnamefont {Shinji}\ \bibnamefont {Yuasa}}, \bibinfo {author}
		{\bibfnamefont {Mark~D.}\ \bibnamefont {Stiles}}, \ and\ \bibinfo {author}
		{\bibfnamefont {Julie}\ \bibnamefont {Grollier}},\ }\bibfield  {title}
	{\enquote {\bibinfo {title} {Neuromorphic computing with nanoscale spintronic
				oscillators},}\ }\href {\doibase 10.1038/nature23011} {\bibfield  {journal}
		{\bibinfo  {journal} {Nature}\ }\textbf {\bibinfo {volume} {547}},\ \bibinfo
		{pages} {428--431} (\bibinfo {year} {2017})}\BibitemShut {NoStop}%
	\bibitem [{\citenamefont {Pickett}\ \emph {et~al.}(2012)\citenamefont
		{Pickett}, \citenamefont {Medeiros-Ribeiro},\ and\ \citenamefont
		{Williams}}]{Pickett12}%
	\BibitemOpen
	\bibfield  {author} {\bibinfo {author} {\bibfnamefont {Matthew~D.}\
			\bibnamefont {Pickett}}, \bibinfo {author} {\bibfnamefont {Gilberto}\
			\bibnamefont {Medeiros-Ribeiro}}, \ and\ \bibinfo {author} {\bibfnamefont
			{R.~Stanley}\ \bibnamefont {Williams}},\ }\bibfield  {title} {\enquote
		{\bibinfo {title} {A scalable neuristor built with mott memristors},}\ }\href
	{\doibase 10.1038/nmat3510} {\bibfield  {journal} {\bibinfo  {journal}
			{Nature Materials}\ }\textbf {\bibinfo {volume} {12}},\ \bibinfo {pages}
		{114--117} (\bibinfo {year} {2012})}\BibitemShut {NoStop}%
	\bibitem [{\citenamefont {Romera}\ \emph {et~al.}(2018)\citenamefont {Romera},
		\citenamefont {Talatchian}, \citenamefont {Tsunegi}, \citenamefont {Araujo},
		\citenamefont {Cros}, \citenamefont {Bortolotti}, \citenamefont {Trastoy},
		\citenamefont {Yakushiji}, \citenamefont {Fukushima}, \citenamefont {Kubota},
		\citenamefont {Yuasa}, \citenamefont {Ernoult}, \citenamefont
		{Vodenicarevic}, \citenamefont {Hirtzlin}, \citenamefont {Locatelli},
		\citenamefont {Querlioz},\ and\ \citenamefont {Grollier}}]{Romera18}%
	\BibitemOpen
	\bibfield  {author} {\bibinfo {author} {\bibfnamefont {Miguel}\ \bibnamefont
			{Romera}}, \bibinfo {author} {\bibfnamefont {Philippe}\ \bibnamefont
			{Talatchian}}, \bibinfo {author} {\bibfnamefont {Sumito}\ \bibnamefont
			{Tsunegi}}, \bibinfo {author} {\bibfnamefont {Flavio~Abreu}\ \bibnamefont
			{Araujo}}, \bibinfo {author} {\bibfnamefont {Vincent}\ \bibnamefont {Cros}},
		\bibinfo {author} {\bibfnamefont {Paolo}\ \bibnamefont {Bortolotti}},
		\bibinfo {author} {\bibfnamefont {Juan}\ \bibnamefont {Trastoy}}, \bibinfo
		{author} {\bibfnamefont {Kay}\ \bibnamefont {Yakushiji}}, \bibinfo {author}
		{\bibfnamefont {Akio}\ \bibnamefont {Fukushima}}, \bibinfo {author}
		{\bibfnamefont {Hitoshi}\ \bibnamefont {Kubota}}, \bibinfo {author}
		{\bibfnamefont {Shinji}\ \bibnamefont {Yuasa}}, \bibinfo {author}
		{\bibfnamefont {Maxence}\ \bibnamefont {Ernoult}}, \bibinfo {author}
		{\bibfnamefont {Damir}\ \bibnamefont {Vodenicarevic}}, \bibinfo {author}
		{\bibfnamefont {Tifenn}\ \bibnamefont {Hirtzlin}}, \bibinfo {author}
		{\bibfnamefont {Nicolas}\ \bibnamefont {Locatelli}}, \bibinfo {author}
		{\bibfnamefont {Damien}\ \bibnamefont {Querlioz}}, \ and\ \bibinfo {author}
		{\bibfnamefont {Julie}\ \bibnamefont {Grollier}},\ }\bibfield  {title}
	{\enquote {\bibinfo {title} {Vowel recognition with four coupled spin-torque
				nano-oscillators},}\ }\href {\doibase 10.1038/s41586-018-0632-y} {\bibfield
		{journal} {\bibinfo  {journal} {Nature}\ }\textbf {\bibinfo {volume} {563}},\
		\bibinfo {pages} {230--234} (\bibinfo {year} {2018})}\BibitemShut {NoStop}%
	\bibitem [{\citenamefont {Markovi{\'{c}}}\ \emph {et~al.}(2019)\citenamefont
		{Markovi{\'{c}}}, \citenamefont {Leroux}, \citenamefont {Riou}, \citenamefont
		{Araujo}, \citenamefont {Torrejon}, \citenamefont {Querlioz}, \citenamefont
		{Fukushima}, \citenamefont {Yuasa}, \citenamefont {Trastoy}, \citenamefont
		{Bortolotti},\ and\ \citenamefont {Grollier}}]{Markovic19}%
	\BibitemOpen
	\bibfield  {author} {\bibinfo {author} {\bibfnamefont {D.}~\bibnamefont
			{Markovi{\'{c}}}}, \bibinfo {author} {\bibfnamefont {N.}~\bibnamefont
			{Leroux}}, \bibinfo {author} {\bibfnamefont {M.}~\bibnamefont {Riou}},
		\bibinfo {author} {\bibfnamefont {F.~Abreu}\ \bibnamefont {Araujo}}, \bibinfo
		{author} {\bibfnamefont {J.}~\bibnamefont {Torrejon}}, \bibinfo {author}
		{\bibfnamefont {D.}~\bibnamefont {Querlioz}}, \bibinfo {author}
		{\bibfnamefont {A.}~\bibnamefont {Fukushima}}, \bibinfo {author}
		{\bibfnamefont {S.}~\bibnamefont {Yuasa}}, \bibinfo {author} {\bibfnamefont
			{J.}~\bibnamefont {Trastoy}}, \bibinfo {author} {\bibfnamefont
			{P.}~\bibnamefont {Bortolotti}}, \ and\ \bibinfo {author} {\bibfnamefont
			{J.}~\bibnamefont {Grollier}},\ }\bibfield  {title} {\enquote {\bibinfo
			{title} {Reservoir computing with the frequency, phase, and amplitude of
				spin-torque nano-oscillators},}\ }\href {\doibase 10.1063/1.5079305}
	{\bibfield  {journal} {\bibinfo  {journal} {Applied Physics Letters}\
		}\textbf {\bibinfo {volume} {114}},\ \bibinfo {pages} {012409} (\bibinfo
		{year} {2019})}\BibitemShut {NoStop}%
	\bibitem [{\citenamefont {Jaeger}(2004)}]{Jaeger04}%
	\BibitemOpen
	\bibfield  {author} {\bibinfo {author} {\bibfnamefont {H.}~\bibnamefont
			{Jaeger}},\ }\bibfield  {title} {\enquote {\bibinfo {title} {Harnessing
				nonlinearity: Predicting chaotic systems and saving energy in wireless
				communication},}\ }\href {\doibase 10.1126/science.1091277} {\bibfield
		{journal} {\bibinfo  {journal} {Science}\ }\textbf {\bibinfo {volume}
			{304}},\ \bibinfo {pages} {78--80} (\bibinfo {year} {2004})}\BibitemShut
	{NoStop}%
	\bibitem [{\citenamefont {Tsunegi}\ \emph {et~al.}(2019)\citenamefont
		{Tsunegi}, \citenamefont {Taniguchi}, \citenamefont {Nakajima}, \citenamefont
		{Miwa}, \citenamefont {Yakushiji}, \citenamefont {Fukushima}, \citenamefont
		{Yuasa},\ and\ \citenamefont {Kubota}}]{Tsunegi19}%
	\BibitemOpen
	\bibfield  {author} {\bibinfo {author} {\bibfnamefont {Sumito}\ \bibnamefont
			{Tsunegi}}, \bibinfo {author} {\bibfnamefont {Tomohiro}\ \bibnamefont
			{Taniguchi}}, \bibinfo {author} {\bibfnamefont {Kohei}\ \bibnamefont
			{Nakajima}}, \bibinfo {author} {\bibfnamefont {Shinji}\ \bibnamefont {Miwa}},
		\bibinfo {author} {\bibfnamefont {Kay}\ \bibnamefont {Yakushiji}}, \bibinfo
		{author} {\bibfnamefont {Akio}\ \bibnamefont {Fukushima}}, \bibinfo {author}
		{\bibfnamefont {Shinji}\ \bibnamefont {Yuasa}}, \ and\ \bibinfo {author}
		{\bibfnamefont {Hitoshi}\ \bibnamefont {Kubota}},\ }\bibfield  {title}
	{\enquote {\bibinfo {title} {Physical reservoir computing based on spin
				torque oscillator with forced synchronization},}\ }\href {\doibase
		10.1063/1.5081797} {\bibfield  {journal} {\bibinfo  {journal} {Applied
				Physics Letters}\ }\textbf {\bibinfo {volume} {114}},\ \bibinfo {pages}
		{164101} (\bibinfo {year} {2019})}\BibitemShut {NoStop}%
	\bibitem [{\citenamefont {Choi}\ and\ \citenamefont {Kim}(2019)}]{Choi19a}%
	\BibitemOpen
	\bibfield  {author} {\bibinfo {author} {\bibfnamefont {Jaesung}\ \bibnamefont
			{Choi}}\ and\ \bibinfo {author} {\bibfnamefont {Pilwon}\ \bibnamefont
			{Kim}},\ }\bibfield  {title} {\enquote {\bibinfo {title} {Critical
				neuromorphic computing based on explosive synchronization},}\ }\href
	{\doibase 10.1063/1.5086902} {\bibfield  {journal} {\bibinfo  {journal}
			{Chaos: An Interdisciplinary Journal of Nonlinear Science}\ }\textbf
		{\bibinfo {volume} {29}},\ \bibinfo {pages} {043110} (\bibinfo {year}
		{2019})}\BibitemShut {NoStop}%
	\bibitem [{\citenamefont {Khymyn}\ \emph {et~al.}(2018)\citenamefont {Khymyn},
		\citenamefont {Lisenkov}, \citenamefont {Voorheis}, \citenamefont
		{Sulymenko}, \citenamefont {Prokopenko}, \citenamefont {Tiberkevich},
		\citenamefont {Akerman},\ and\ \citenamefont {Slavin}}]{Khymyn18}%
	\BibitemOpen
	\bibfield  {author} {\bibinfo {author} {\bibfnamefont {Roman}\ \bibnamefont
			{Khymyn}}, \bibinfo {author} {\bibfnamefont {Ivan}\ \bibnamefont {Lisenkov}},
		\bibinfo {author} {\bibfnamefont {James}\ \bibnamefont {Voorheis}}, \bibinfo
		{author} {\bibfnamefont {Olga}\ \bibnamefont {Sulymenko}}, \bibinfo {author}
		{\bibfnamefont {Oleksandr}\ \bibnamefont {Prokopenko}}, \bibinfo {author}
		{\bibfnamefont {Vasil}\ \bibnamefont {Tiberkevich}}, \bibinfo {author}
		{\bibfnamefont {Johan}\ \bibnamefont {Akerman}}, \ and\ \bibinfo {author}
		{\bibfnamefont {Andrei}\ \bibnamefont {Slavin}},\ }\bibfield  {title}
	{\enquote {\bibinfo {title} {Ultra-fast artificial neuron: generation of
				picosecond-duration spikes in a current-driven antiferromagnetic
				auto-oscillator},}\ }\href {\doibase 10.1038/s41598-018-33697-0} {\bibfield
		{journal} {\bibinfo  {journal} {Scientific Reports}\ }\textbf {\bibinfo
			{volume} {8}},\ \bibinfo {pages} {15727} (\bibinfo {year}
		{2018})}\BibitemShut {NoStop}%
	\bibitem [{\citenamefont {Zhao}\ \emph {et~al.}(2015)\citenamefont {Zhao},
		\citenamefont {Danesh}, \citenamefont {Wysocki},\ and\ \citenamefont
		{Yi}}]{Zhao15}%
	\BibitemOpen
	\bibfield  {author} {\bibinfo {author} {\bibfnamefont {Chenyuan}\
			\bibnamefont {Zhao}}, \bibinfo {author} {\bibfnamefont {Wafi}\ \bibnamefont
			{Danesh}}, \bibinfo {author} {\bibfnamefont {Bryant~T.}\ \bibnamefont
			{Wysocki}}, \ and\ \bibinfo {author} {\bibfnamefont {Yang}\ \bibnamefont
			{Yi}},\ }\bibfield  {title} {\enquote {\bibinfo {title} {Neuromorphic
				encoding system design with chaos based {CMOS} analog neuron},}\ }in\ \href
	{\doibase 10.1109/cisda.2015.7208631} {\emph {\bibinfo {booktitle} {2015
				{IEEE} Symposium on Computational Intelligence for Security and Defense
				Applications ({CISDA})}}}\ (\bibinfo  {publisher} {{IEEE}},\ \bibinfo {year}
	{2015})\BibitemShut {NoStop}%
	\bibitem [{\citenamefont {Matsumoto}\ \emph {et~al.}(2019)\citenamefont
		{Matsumoto}, \citenamefont {Lequeux}, \citenamefont {Imamura},\ and\
		\citenamefont {Grollier}}]{Matsumoto19}%
	\BibitemOpen
	\bibfield  {author} {\bibinfo {author} {\bibfnamefont {R.}~\bibnamefont
			{Matsumoto}}, \bibinfo {author} {\bibfnamefont {S.}~\bibnamefont {Lequeux}},
		\bibinfo {author} {\bibfnamefont {H.}~\bibnamefont {Imamura}}, \ and\
		\bibinfo {author} {\bibfnamefont {J.}~\bibnamefont {Grollier}},\ }\bibfield
	{title} {\enquote {\bibinfo {title} {Chaos and relaxation oscillations in
				spin-torque windmill spiking oscillators},}\ }\href {\doibase
		10.1103/physrevapplied.11.044093} {\bibfield  {journal} {\bibinfo  {journal}
			{Physical Review Applied}\ }\textbf {\bibinfo {volume} {11}},\ \bibinfo
		{pages} {044093} (\bibinfo {year} {2019})}\BibitemShut {NoStop}%
	\bibitem [{\citenamefont {Gomonay}\ and\ \citenamefont
		{Loktev}(2010)}]{Gomonay10}%
	\BibitemOpen
	\bibfield  {author} {\bibinfo {author} {\bibfnamefont {Helen~V.}\
			\bibnamefont {Gomonay}}\ and\ \bibinfo {author} {\bibfnamefont {Vadim~M.}\
			\bibnamefont {Loktev}},\ }\bibfield  {title} {\enquote {\bibinfo {title}
			{Spin transfer and current-induced switching in antiferromagnets},}\ }\href
	{\doibase 10.1103/PhysRevB.81.144427} {\bibfield  {journal} {\bibinfo
			{journal} {Physical Review B}\ }\textbf {\bibinfo {volume} {81}},\ \bibinfo
		{pages} {144427} (\bibinfo {year} {2010})}\BibitemShut {NoStop}%
	\bibitem [{\citenamefont {Cheng}\ \emph {et~al.}(2016)\citenamefont {Cheng},
		\citenamefont {Xiao},\ and\ \citenamefont {Brataas}}]{Cheng16a}%
	\BibitemOpen
	\bibfield  {author} {\bibinfo {author} {\bibfnamefont {Ran}\ \bibnamefont
			{Cheng}}, \bibinfo {author} {\bibfnamefont {Di}~\bibnamefont {Xiao}}, \ and\
		\bibinfo {author} {\bibfnamefont {Arne}\ \bibnamefont {Brataas}},\ }\bibfield
	{title} {\enquote {\bibinfo {title} {Terahertz antiferromagnetic spin hall
				nano-oscillator},}\ }\href {\doibase 10.1103/physrevlett.116.207603}
	{\bibfield  {journal} {\bibinfo  {journal} {Physical Review Letters}\
		}\textbf {\bibinfo {volume} {116}},\ \bibinfo {pages} {207603} (\bibinfo
		{year} {2016})}\BibitemShut {NoStop}%
	\bibitem [{\citenamefont {Parthasarathy}\ \emph {et~al.}()\citenamefont
		{Parthasarathy}, \citenamefont {Cogulu}, \citenamefont {Kent},\ and\
		\citenamefont {Rakheja}}]{Parthasarathy19a}%
	\BibitemOpen
	\bibfield  {author} {\bibinfo {author} {\bibfnamefont {Arun}\ \bibnamefont
			{Parthasarathy}}, \bibinfo {author} {\bibfnamefont {Egecan}\ \bibnamefont
			{Cogulu}}, \bibinfo {author} {\bibfnamefont {Andrew~D.}\ \bibnamefont
			{Kent}}, \ and\ \bibinfo {author} {\bibfnamefont {Shaloo}\ \bibnamefont
			{Rakheja}},\ }\bibfield  {title} {\enquote {\bibinfo {title} {Precessional
				spin-torque dynamics in biaxial antiferromagnets},}\ }\href@noop {} {\
	}\Eprint {http://arxiv.org/abs/1911.00445} {1911.00445} \BibitemShut
	{NoStop}%
	\bibitem [{\citenamefont {Locatelli}\ \emph {et~al.}(2013)\citenamefont
		{Locatelli}, \citenamefont {Cros},\ and\ \citenamefont
		{Grollier}}]{Locatelli13a}%
	\BibitemOpen
	\bibfield  {author} {\bibinfo {author} {\bibfnamefont {N.}~\bibnamefont
			{Locatelli}}, \bibinfo {author} {\bibfnamefont {V.}~\bibnamefont {Cros}}, \
		and\ \bibinfo {author} {\bibfnamefont {J.}~\bibnamefont {Grollier}},\
	}\bibfield  {title} {\enquote {\bibinfo {title} {Spin-torque building
				blocks},}\ }\href {\doibase 10.1038/nmat3823} {\bibfield  {journal} {\bibinfo
			{journal} {Nature Materials}\ }\textbf {\bibinfo {volume} {13}},\ \bibinfo
		{pages} {11--20} (\bibinfo {year} {2013})}\BibitemShut {NoStop}%
	\bibitem [{\citenamefont {Chen}\ \emph {et~al.}(2016)\citenamefont {Chen},
		\citenamefont {Dumas}, \citenamefont {Eklund}, \citenamefont {Muduli},
		\citenamefont {Houshang}, \citenamefont {Awad}, \citenamefont {Durrenfeld},
		\citenamefont {Malm}, \citenamefont {Rusu},\ and\ \citenamefont
		{Akerman}}]{Chen16a}%
	\BibitemOpen
	\bibfield  {author} {\bibinfo {author} {\bibfnamefont {Tingsu}\ \bibnamefont
			{Chen}}, \bibinfo {author} {\bibfnamefont {Randy~K.}\ \bibnamefont {Dumas}},
		\bibinfo {author} {\bibfnamefont {Anders}\ \bibnamefont {Eklund}}, \bibinfo
		{author} {\bibfnamefont {Pranaba~K.}\ \bibnamefont {Muduli}}, \bibinfo
		{author} {\bibfnamefont {Afshin}\ \bibnamefont {Houshang}}, \bibinfo {author}
		{\bibfnamefont {Ahmad~A.}\ \bibnamefont {Awad}}, \bibinfo {author}
		{\bibfnamefont {Philipp}\ \bibnamefont {Durrenfeld}}, \bibinfo {author}
		{\bibfnamefont {B.~Gunnar}\ \bibnamefont {Malm}}, \bibinfo {author}
		{\bibfnamefont {Ana}\ \bibnamefont {Rusu}}, \ and\ \bibinfo {author}
		{\bibfnamefont {Johan}\ \bibnamefont {Akerman}},\ }\bibfield  {title}
	{\enquote {\bibinfo {title} {Spin-torque and spin-hall nano-oscillators},}\
	}\href {\doibase 10.1109/jproc.2016.2554518} {\bibfield  {journal} {\bibinfo
			{journal} {Proceedings of the {IEEE}}\ }\textbf {\bibinfo {volume} {104}},\
		\bibinfo {pages} {1919--1945} (\bibinfo {year} {2016})}\BibitemShut {NoStop}%
	\bibitem [{\citenamefont {Slonczewski}(1996)}]{Slonczewski96}%
	\BibitemOpen
	\bibfield  {author} {\bibinfo {author} {\bibfnamefont {J.~C.}\ \bibnamefont
			{Slonczewski}},\ }\bibfield  {title} {\enquote {\bibinfo {title}
			{Current-driven excitation of magnetic multilayers},}\ }\href
	{http://dx.doi.org/10.1016/0304-8853(96)00062-5} {\bibfield  {journal}
		{\bibinfo  {journal} {Journal of Magnetism and Magnetic Materials}\ }\textbf
		{\bibinfo {volume} {159}},\ \bibinfo {pages} {L1--L7} (\bibinfo {year}
		{1996})}\BibitemShut {NoStop}%
	\bibitem [{\citenamefont {Slonczewski}(2002)}]{Slonczewski02}%
	\BibitemOpen
	\bibfield  {author} {\bibinfo {author} {\bibfnamefont {J.~C.}\ \bibnamefont
			{Slonczewski}},\ }\bibfield  {title} {\enquote {\bibinfo {title} {Currents
				and torques in metallic magnetic multilayers},}\ }\href
	{http://www.sciencedirect.com/science/article/B6TJJ-45RFP8K-5/2/74b7bbf5e7dc831281ad65874ccb88b9}
	{\bibfield  {journal} {\bibinfo  {journal} {Journal of Magnetism and Magnetic
				Materials}\ }\textbf {\bibinfo {volume} {247}},\ \bibinfo {pages} {324--338}
		(\bibinfo {year} {2002})}\BibitemShut {NoStop}%
	\bibitem [{\citenamefont {Xiao}\ \emph {et~al.}()\citenamefont {Xiao},
		\citenamefont {Zangwill},\ and\ \citenamefont {Stiles}}]{Xiao04}%
	\BibitemOpen
	\bibfield  {author} {\bibinfo {author} {\bibfnamefont {Jiang}\ \bibnamefont
			{Xiao}}, \bibinfo {author} {\bibfnamefont {A.}~\bibnamefont {Zangwill}}, \
		and\ \bibinfo {author} {\bibfnamefont {M.~D.}\ \bibnamefont {Stiles}},\
	}\bibfield  {title} {\enquote {\bibinfo {title} {Boltzmann test of
				slonczewski's theory of spin-transfer torque},}\ }\href {\doibase
		10.1103/physrevb.70.172405} {\ \textbf {\bibinfo {volume} {70}},\ \bibinfo
		{pages} {172405}}\BibitemShut {NoStop}%
	\bibitem [{\citenamefont {Baryakhtar}\ and\ \citenamefont
		{Ivanov}(1980)}]{Baryakhtar80}%
	\BibitemOpen
	\bibfield  {author} {\bibinfo {author} {\bibfnamefont {I.~V.}\ \bibnamefont
			{Baryakhtar}}\ and\ \bibinfo {author} {\bibfnamefont {B.~A.}\ \bibnamefont
			{Ivanov}},\ }\bibfield  {title} {\enquote {\bibinfo {title} {Nonlinear waves
				in antiferromagnets},}\ }\href {\doibase 10.1016/0038-1098(80)90148-9}
	{\bibfield  {journal} {\bibinfo  {journal} {Solid State Communications}\
		}\textbf {\bibinfo {volume} {34}},\ \bibinfo {pages} {545--547} (\bibinfo
		{year} {1980})}\BibitemShut {NoStop}%
	\bibitem [{\citenamefont {Akhiezer}\ \emph {et~al.}(1968)\citenamefont
		{Akhiezer}, \citenamefont {Bar'yakhtar},\ and\ \citenamefont
		{Peletminski\u{\i}}}]{Akhiezer68}%
	\BibitemOpen
	\bibfield  {author} {\bibinfo {author} {\bibfnamefont {A.~I.}\ \bibnamefont
			{Akhiezer}}, \bibinfo {author} {\bibfnamefont {V.~G.}\ \bibnamefont
			{Bar'yakhtar}}, \ and\ \bibinfo {author} {\bibfnamefont {S.~V.}\ \bibnamefont
			{Peletminski\u{\i}}},\ }\href@noop {} {\emph {\bibinfo {title} {Spin
				waves}}},\ edited by\ \bibinfo {editor} {\bibfnamefont {G.}~\bibnamefont
		{H{\"o}hler}}\ (\bibinfo  {publisher} {North--Holland},\ \bibinfo {address}
	{Amsterdam},\ \bibinfo {year} {1968})\BibitemShut {NoStop}%
	\bibitem [{\citenamefont {Arkady~Pikovsky}(2016)}]{ArkadyPikovsky16}%
	\BibitemOpen
	\bibfield  {author} {\bibinfo {author} {\bibfnamefont {Antonio~Politi}\
			\bibnamefont {Arkady~Pikovsky}},\ }\href
	{https://www.ebook.de/de/product/24385701/arkady_pikovsky_antonio_politi_lyapunov_exponents_a_tool_to_explore_complex_dynamics.html}
	{\emph {\bibinfo {title} {Lyapunov Exponents: A Tool to Explore Complex
				Dynamics}}}\ (\bibinfo  {publisher} {CAMBRIDGE},\ \bibinfo {year}
	{2016})\BibitemShut {NoStop}%
	\bibitem [{\citenamefont {Cencini}\ \emph {et~al.}(2009)\citenamefont
		{Cencini}, \citenamefont {Cecconi},\ and\ \citenamefont
		{Vulpiani}}]{Cencini09}%
	\BibitemOpen
	\bibfield  {author} {\bibinfo {author} {\bibfnamefont {Massimo}\ \bibnamefont
			{Cencini}}, \bibinfo {author} {\bibfnamefont {Fabio}\ \bibnamefont
			{Cecconi}}, \ and\ \bibinfo {author} {\bibfnamefont {Angelo}\ \bibnamefont
			{Vulpiani}},\ }\href
	{https://www.ebook.de/de/product/9115123/massimo_cencini_fabio_cecconi_angelo_vulpiani_chaos_from_simple_models_to_complex_systems.html}
	{\emph {\bibinfo {title} {Chaos: From Simple Models to Complex Systems}}}\
	(\bibinfo  {publisher} {WORLD SCIENTIFIC PUB CO INC},\ \bibinfo {year}
	{2009})\BibitemShut {NoStop}%
	\bibitem [{Note1()}]{Note1}%
	\BibitemOpen
	\bibinfo {note} {The Lyapunov exponents have an intuitive geometrical
		meaning: under the time evolution an infinitesimally small $d$-dimensional
		sphere in the phase space is deforming into an ellipsoid with semi-axes
		$\propto e^{\Lambda _i\tau }$.}\BibitemShut {Stop}%
	\bibitem [{\citenamefont {Oseledets}(1968)}]{Oseledets68}%
	\BibitemOpen
	\bibfield  {author} {\bibinfo {author} {\bibfnamefont {V.~I.}\ \bibnamefont
			{Oseledets}},\ }\bibfield  {title} {\enquote {\bibinfo {title} {A
				multiplicative ergodic theorem. lyapunov characteristic numbers for dynamical
				systems.}}\ }\href@noop {} {\bibfield  {journal} {\bibinfo  {journal} {Trans.
				Mosc. Math. Soc.}\ }\textbf {\bibinfo {volume} {19}},\ \bibinfo {pages} {179}
		(\bibinfo {year} {1968})}\BibitemShut {NoStop}%
	\bibitem [{\citenamefont {Benettin}\ \emph
		{et~al.}(1980{\natexlab{a}})\citenamefont {Benettin}, \citenamefont
		{Galgani}, \citenamefont {Giorgilli},\ and\ \citenamefont
		{Strelcyn}}]{Benettin80}%
	\BibitemOpen
	\bibfield  {author} {\bibinfo {author} {\bibfnamefont {Giancarlo}\
			\bibnamefont {Benettin}}, \bibinfo {author} {\bibfnamefont {Luigi}\
			\bibnamefont {Galgani}}, \bibinfo {author} {\bibfnamefont {Antonio}\
			\bibnamefont {Giorgilli}}, \ and\ \bibinfo {author} {\bibfnamefont
			{Jean-Marie}\ \bibnamefont {Strelcyn}},\ }\bibfield  {title} {\enquote
		{\bibinfo {title} {Lyapunov characteristic exponents for smooth dynamical
				systems and for hamiltonian systems: a method for computing all of them. part
				1: Theory},}\ }\href {\doibase 10.1007/bf02128236} {\bibfield  {journal}
		{\bibinfo  {journal} {Meccanica}\ }\textbf {\bibinfo {volume} {15}},\
		\bibinfo {pages} {9--20} (\bibinfo {year} {1980}{\natexlab{a}})}\BibitemShut
	{NoStop}%
	\bibitem [{\citenamefont {Benettin}\ \emph
		{et~al.}(1980{\natexlab{b}})\citenamefont {Benettin}, \citenamefont
		{Galgani}, \citenamefont {Giorgilli},\ and\ \citenamefont
		{Strelcyn}}]{Benettin80a}%
	\BibitemOpen
	\bibfield  {author} {\bibinfo {author} {\bibfnamefont {Giancarlo}\
			\bibnamefont {Benettin}}, \bibinfo {author} {\bibfnamefont {Luigi}\
			\bibnamefont {Galgani}}, \bibinfo {author} {\bibfnamefont {Antonio}\
			\bibnamefont {Giorgilli}}, \ and\ \bibinfo {author} {\bibfnamefont
			{Jean-Marie}\ \bibnamefont {Strelcyn}},\ }\bibfield  {title} {\enquote
		{\bibinfo {title} {Lyapunov characteristic exponents for smooth dynamical
				systems and for hamiltonian systems: A method for computing all of them. part
				2: Numerical application},}\ }\href {\doibase 10.1007/bf02128237} {\bibfield
		{journal} {\bibinfo  {journal} {Meccanica}\ }\textbf {\bibinfo {volume}
			{15}},\ \bibinfo {pages} {21--30} (\bibinfo {year}
		{1980}{\natexlab{b}})}\BibitemShut {NoStop}%
	\bibitem [{Note2()}]{Note2}%
	\BibitemOpen
	\bibinfo {note} {The statement $S_d=-2\protect \mathaccentV {bar}016{\alpha
		}$ was established numerically for a large range of the parameters $b$, $j$,
		$\protect \mathaccentV {bar}016{\alpha }$ and initial
		conditions.}\BibitemShut {Stop}%
	\bibitem [{\citenamefont {Ott}()}]{Ott93}%
	\BibitemOpen
	\bibfield  {author} {\bibinfo {author} {\bibfnamefont {Edward}\ \bibnamefont
			{Ott}},\ }\href@noop {} {\emph {\bibinfo {title} {Chaos in dynamical
				systems}}}\ (\bibinfo  {publisher} {Cambridge University Press})\BibitemShut
	{NoStop}%
	\bibitem [{\citenamefont {Stankevich}\ \emph {et~al.}(2019)\citenamefont
		{Stankevich}, \citenamefont {Kuznetsov}, \citenamefont {Popova},\ and\
		\citenamefont {Seleznev}}]{Stankevich19}%
	\BibitemOpen
	\bibfield  {author} {\bibinfo {author} {\bibfnamefont {Nataliya}\
			\bibnamefont {Stankevich}}, \bibinfo {author} {\bibfnamefont {Alexander}\
			\bibnamefont {Kuznetsov}}, \bibinfo {author} {\bibfnamefont {Elena}\
			\bibnamefont {Popova}}, \ and\ \bibinfo {author} {\bibfnamefont {Evgeniy}\
			\bibnamefont {Seleznev}},\ }\bibfield  {title} {\enquote {\bibinfo {title}
			{Chaos and hyperchaos via secondary neimark{\textendash}sacker bifurcation in
				a model of radiophysical generator},}\ }\href {\doibase
		10.1007/s11071-019-05132-0} {\bibfield  {journal} {\bibinfo  {journal}
			{Nonlinear Dynamics}\ }\textbf {\bibinfo {volume} {97}},\ \bibinfo {pages}
		{2355--2370} (\bibinfo {year} {2019})}\BibitemShut {NoStop}%
	\bibitem [{\citenamefont {Vitolo}\ \emph {et~al.}(2011)\citenamefont {Vitolo},
		\citenamefont {Broer},\ and\ \citenamefont {Sim{\'{o}}}}]{Vitolo11}%
	\BibitemOpen
	\bibfield  {author} {\bibinfo {author} {\bibfnamefont {Renato}\ \bibnamefont
			{Vitolo}}, \bibinfo {author} {\bibfnamefont {Henk}\ \bibnamefont {Broer}}, \
		and\ \bibinfo {author} {\bibfnamefont {Carles}\ \bibnamefont {Sim{\'{o}}}},\
	}\bibfield  {title} {\enquote {\bibinfo {title} {Quasi-periodic bifurcations
				of invariant circles in low-dimensional dissipative dynamical systems},}\
	}\href {\doibase 10.1134/s1560354711010060} {\bibfield  {journal} {\bibinfo
			{journal} {Regular and Chaotic Dynamics}\ }\textbf {\bibinfo {volume} {16}},\
		\bibinfo {pages} {154--184} (\bibinfo {year} {2011})}\BibitemShut {NoStop}%
	\bibitem [{\citenamefont {Kuznetsov}\ \emph {et~al.}(2015)\citenamefont
		{Kuznetsov}, \citenamefont {Migunova}, \citenamefont {Sataev}, \citenamefont
		{Sedova},\ and\ \citenamefont {Turukina}}]{Kuznetsov15}%
	\BibitemOpen
	\bibfield  {author} {\bibinfo {author} {\bibfnamefont {Alexander~P.}\
			\bibnamefont {Kuznetsov}}, \bibinfo {author} {\bibfnamefont {Natalia~A.}\
			\bibnamefont {Migunova}}, \bibinfo {author} {\bibfnamefont {Igor~R.}\
			\bibnamefont {Sataev}}, \bibinfo {author} {\bibfnamefont {Yuliya~V.}\
			\bibnamefont {Sedova}}, \ and\ \bibinfo {author} {\bibfnamefont {Ludmila~V.}\
			\bibnamefont {Turukina}},\ }\bibfield  {title} {\enquote {\bibinfo {title}
			{From chaos to quasi-periodicity},}\ }\href {\doibase
		10.1134/s1560354715020070} {\bibfield  {journal} {\bibinfo  {journal}
			{Regular and Chaotic Dynamics}\ }\textbf {\bibinfo {volume} {20}},\ \bibinfo
		{pages} {189--204} (\bibinfo {year} {2015})}\BibitemShut {NoStop}%
	\bibitem [{\citenamefont {Bar'yakhtar}\ and\ \citenamefont
		{Ivanov}(1979)}]{Baryakhtar79}%
	\BibitemOpen
	\bibfield  {author} {\bibinfo {author} {\bibfnamefont {I.~V.}\ \bibnamefont
			{Bar'yakhtar}}\ and\ \bibinfo {author} {\bibfnamefont {B.~A.}\ \bibnamefont
			{Ivanov}},\ }\bibfield  {title} {\enquote {\bibinfo {title} {Nonlinear
				magnetization waves in the antiferromagnet},}\ }\href@noop {} {\bibfield
		{journal} {\bibinfo  {journal} {Sov. J. Low Temp. Phys.}\ }\textbf {\bibinfo
			{volume} {5}},\ \bibinfo {pages} {2620} (\bibinfo {year} {1979})}\BibitemShut
	{NoStop}%
	\bibitem [{\citenamefont {Ivanov}\ and\ \citenamefont
		{Kolezhuk}(1995)}]{Ivanov95e}%
	\BibitemOpen
	\bibfield  {author} {\bibinfo {author} {\bibfnamefont {B.~A.}\ \bibnamefont
			{Ivanov}}\ and\ \bibinfo {author} {\bibfnamefont {A.~K.}\ \bibnamefont
			{Kolezhuk}},\ }\bibfield  {title} {\enquote {\bibinfo {title} {Solitons in
				low-dimensional antiferromagnets},}\ }\href {\doibase 10.1063/1.593081}
	{\bibfield  {journal} {\bibinfo  {journal} {Low Temperature Physics}\
		}\textbf {\bibinfo {volume} {21}},\ \bibinfo {pages} {275--301} (\bibinfo
		{year} {1995})}\BibitemShut {NoStop}%
	\bibitem [{\citenamefont {Turov}\ \emph {et~al.}(2001)\citenamefont {Turov},
		\citenamefont {Kolchanov}, \citenamefont {Menshenin}, \citenamefont
		{Mirsayev},\ and\ \citenamefont {Nikolaev}}]{Turov01en}%
	\BibitemOpen
	\bibfield  {author} {\bibinfo {author} {\bibfnamefont {E.~A.}\ \bibnamefont
			{Turov}}, \bibinfo {author} {\bibfnamefont {A.~V.}\ \bibnamefont
			{Kolchanov}}, \bibinfo {author} {\bibfnamefont {V.~V.}\ \bibnamefont
			{Menshenin}}, \bibinfo {author} {\bibfnamefont {I.~F.}\ \bibnamefont
			{Mirsayev}}, \ and\ \bibinfo {author} {\bibfnamefont {V.~V.}\ \bibnamefont
			{Nikolaev}},\ }\href@noop {} {\emph {\bibinfo {title} {Symmetry and physical
				properties of antiferromagnets}}}\ (\bibinfo  {publisher} {FIZMATLIT},\
	\bibinfo {address} {Moscow},\ \bibinfo {year} {2001})\BibitemShut {NoStop}%
	\bibitem [{\citenamefont {Kaganov}\ and\ \citenamefont
		{Tsukernik}(1958)}]{Kaganov58}%
	\BibitemOpen
	\bibfield  {author} {\bibinfo {author} {\bibfnamefont {M.~I.}\ \bibnamefont
			{Kaganov}}\ and\ \bibinfo {author} {\bibfnamefont {V.~M.}\ \bibnamefont
			{Tsukernik}},\ }\bibfield  {title} {\enquote {\bibinfo {title} {Contribution
				to the theory of antiferromagnetism at low temperatures},}\ }\href
	{http://www.jetp.ac.ru/cgi-bin/e/index/r/34/1/p106?a=list} {\bibfield
		{journal} {\bibinfo  {journal} {Soviet Physics JETP}\ }\textbf {\bibinfo
			{volume} {34(7)}},\ \bibinfo {pages} {73} (\bibinfo {year}
		{1958})}\BibitemShut {NoStop}%
	\bibitem [{\citenamefont {Mascolo}(2019)}]{Mascolo19}%
	\BibitemOpen
	\bibfield  {author} {\bibinfo {author} {\bibfnamefont {Ida}\ \bibnamefont
			{Mascolo}},\ }\bibfield  {title} {\enquote {\bibinfo {title} {Recent
				developments in the dynamic stability of elastic structures},}\ }\href
	{\doibase 10.3389/fams.2019.00051} {\bibfield  {journal} {\bibinfo  {journal}
			{Frontiers in Applied Mathematics and Statistics}\ }\textbf {\bibinfo
			{volume} {5}} (\bibinfo {year} {2019}),\ 10.3389/fams.2019.00051}\BibitemShut
	{NoStop}%
	\bibitem [{\citenamefont {Ziegler}(1952)}]{Ziegler52}%
	\BibitemOpen
	\bibfield  {author} {\bibinfo {author} {\bibfnamefont {H.}~\bibnamefont
			{Ziegler}},\ }\bibfield  {title} {\enquote {\bibinfo {title} {Die
				stabilit{\"a}tskriterien der elastomechanik},}\ }\href {\doibase
		10.1007/bf00536796} {\bibfield  {journal} {\bibinfo  {journal}
			{Ingenieur-Archiv}\ }\textbf {\bibinfo {volume} {20}},\ \bibinfo {pages}
		{49--56} (\bibinfo {year} {1952})}\BibitemShut {NoStop}%
	\bibitem [{\citenamefont {Argyris}\ \emph {et~al.}(2015)\citenamefont
		{Argyris}, \citenamefont {Faust}, \citenamefont {Haase},\ and\ \citenamefont
		{Friedrich}}]{Argyris15}%
	\BibitemOpen
	\bibfield  {author} {\bibinfo {author} {\bibfnamefont {John~H.}\ \bibnamefont
			{Argyris}}, \bibinfo {author} {\bibfnamefont {Gunter}\ \bibnamefont {Faust}},
		\bibinfo {author} {\bibfnamefont {Maria}\ \bibnamefont {Haase}}, \ and\
		\bibinfo {author} {\bibfnamefont {Rudolf}\ \bibnamefont {Friedrich}},\ }\href
	{\doibase 10.1007/978-3-662-46042-9} {\emph {\bibinfo {title} {An Exploration
				of Dynamical Systems and Chaos}}}\ (\bibinfo  {publisher} {Springer Berlin
		Heidelberg},\ \bibinfo {year} {2015})\BibitemShut {NoStop}%
\end{thebibliography}

%

\end{document}